\documentclass[aps,prc,onecolumn,nofootinbib,superscriptaddress]{revtex4-2}
\usepackage{amsmath,amssymb}
\usepackage{graphicx}
\usepackage{dcolumn}
\usepackage{bm}
\usepackage{hyperref}
\usepackage{subcaption}
\usepackage{placeins}
\usepackage{booktabs}
\usepackage{array}
\usepackage{xcolor}
\hypersetup{hidelinks}
\newcommand{\red}[1]{#1}

\newcommand{\blu}[1]{#1}

\begin{document}

\title{Phenomenological Criteria of Halo Nuclei in Ne Isotopes via Diffuseness and Helm Model Approaches with Reaction Cross Sections}

\author{Heesung Kwon}
\affiliation{Department of Physics, Soongsil University, Seoul 06978, Korea}
\author{Kyoungsu Heo}
\email{pleasewhy@ssu.ac.kr}
\affiliation{Department of Physics, Soongsil University, Seoul 06978, Korea}
\author{Seonghyun Kim}
\affiliation{Department of Physics, Soongsil University, Seoul 06978, Korea}
\author{Eunja Ha}
\affiliation{Department of Physics and Research Institute for Natural Science,
Hanyang University, Seoul 04763, Korea}
\author{Myung-Ki Cheoun}
\affiliation{Department of Physics, Soongsil University, Seoul 06978, Korea}

\begin{abstract}

We present a systematic study of halo characteristics in the
neutron-rich isotopes $^{28\text{--}32}$Ne within the
deformed relativistic Hartree--Bogoliubov theory in continuum (DRHBc).
Microscopic density distributions are analyzed in coordinate space,
momentum space, and reaction observables to establish
a quantitative and locally defined criterion for halo identification
in medium-mass nuclei.

The DRHBc densities reveal a pronounced neutron extension
in $^{31}$Ne.
A phenomenological analysis based on deformed Woods--Saxon
fits shows a clear isotopic anomaly in the surface diffuseness
parameter, with $a \approx 1.1$ fm for $^{31}$Ne,
significantly larger than those of neighboring isotopes.
\red{The anomalously large diffuseness is therefore treated as the primary phenomenological halo signature, whereas the reduced fitted radius parameter is used only as a supporting consequence of the chosen normalization and tail-sensitive fit.}
Helm-model form-factor analysis demonstrates that deformation
contributes to geometric smearing but does not fully account
for the extended spatial structure, as reflected in the enhanced
difference between microscopic and folded rms radii.
Glauber reaction cross section calculations further confirm
\red{a robust relative enhancement of the interaction cross section for
$^{31}$Ne that persists across reasonable nucleon--nucleon
interaction prescriptions.}

These complementary analyses consistently identify
$^{31}$Ne as the most prominent halo candidate within
the $^{28\text{--}32}$Ne isotopic chain,
while $^{32}$Ne exhibits intermediate features
and $^{29}$Ne shows no clear halo signature.
The present framework provides a practical and quantitative
approach for identifying halo phenomena in deformed,
neutron-rich nuclei beyond the light-mass region.


\end{abstract}

\maketitle

\section{Introduction}

Nuclei close to the particle drip lines exhibit a variety of exotic structural phenomena that challenge the conventional picture of finite many--body systems. Among them, halo nuclei are characterized by weakly bound valence nucleons that extend far beyond a compact core, producing long-range density tails and enhanced reaction cross sections \cite{Tanihata1985,Hansen1995,Jonson2004}. Since the first observation of halo signatures through the enhanced interaction cross section of $^{11}$Li \cite{Tanihata1985}, halo phenomena have been firmly established in several light nuclei and extensively studied from both experimental and theoretical perspectives \cite{Tanihata2013}.

\red{In medium--mass and heavier nuclei, however, identifying halo structures is considerably more subtle because deformation, pairing correlations, and continuum coupling can mimic or obscure the signatures of spatial extension \cite{Bertulani2007,Hamamoto2010}.} The relative contribution of one or two valence nucleons to the total density distribution is reduced, deformation effects become significant, and conventional indicators such as the root--mean--square (rms) radius may lose their discriminating power. As a result, existing halo criteria are often qualitative or model dependent, and a unified and quantitative framework applicable to medium--mass nuclei is still lacking.

Neon isotopes provide an ideal testing ground for this problem. Located near the so--called island of inversion, neutron--rich Ne isotopes exhibit strong shell evolution, deformation, and weak binding effects. \red{In particular, $^{31}$Ne has long been discussed as a candidate for a deformed one-neutron halo nucleus based on Coulomb breakup and interaction cross-section measurements \cite{Nakamura2009,Takechi2012}.} \red{The deformed halo interpretation of $^{31}$Ne has also been examined in particle-rotor and Glauber-model studies, which showed that weak binding and deformation jointly govern its Coulomb-breakup and reaction observables \cite{Urata2011,Urata2012}.} Neighboring isotopes such as $^{29}$Ne and $^{32}$Ne remain controversial. This situation raises a fundamental question: how can halo structures be identified in a consistent and quantitative manner when deformation and continuum effects coexist? \red{This motivates a strategy based on multiple complementary diagnostics combining microscopic densities, phenomenological parametrizations, and reaction observables.}

The novelty of the present work is not to re-establish the halo character of $^{31}$Ne itself, but to formulate a quantitative isotopic-chain-based diagnostic that combines coordinate-space, form-factor, and reaction observables in a unified way.

In this work, we address this question by performing a systematic analysis of $^{28\text{--}32}$Ne isotopes. We first examine microscopic density distributions obtained from the deformed relativistic Hartree--Bogoliubov theory in continuum (DRHBc), originally formulated in Refs.~\cite{Zhou2010,Li2012} and recently extended to large-scale applications in Ref.~\cite{DRHBc2024}, and analyze their spatial characteristics in coordinate space. Building on these results, we compare existing global halo indicators with locally defined, phenomenological criteria, and further investigate how extended density tails manifest themselves in momentum--space observables \blu{and also in the interaction cross section.} The present paper is organized as follows. In Sec.~II, we analyze the DRHBc density distributions and identify structural anomalies associated with halo formation. Sections~III and IV are devoted to the development of phenomenological criteria and form--factor analyses, respectively. \red{Sec.~V discusses the reaction cross sections as an indicator of halo nuclei. Finally, a combined summary and outlook are presented in Sec.~VI.}

\section{Structural diagnosis based on DRHBc density distributions}

The purpose of this section is to identify structural signatures of halo formation directly from microscopic nuclear density distributions, without invoking reaction observables or phenomenological fitting procedures. At this stage, we do not attempt to define a halo nucleus in a strict sense; instead, we aim to determine which Ne isotopes exhibit anomalous spatial extensions that warrant further quantitative analysis in the subsequent sections.

\subsection{DRHBc framework and deformation properties}

\begin{figure}
\includegraphics[width=0.45\linewidth]{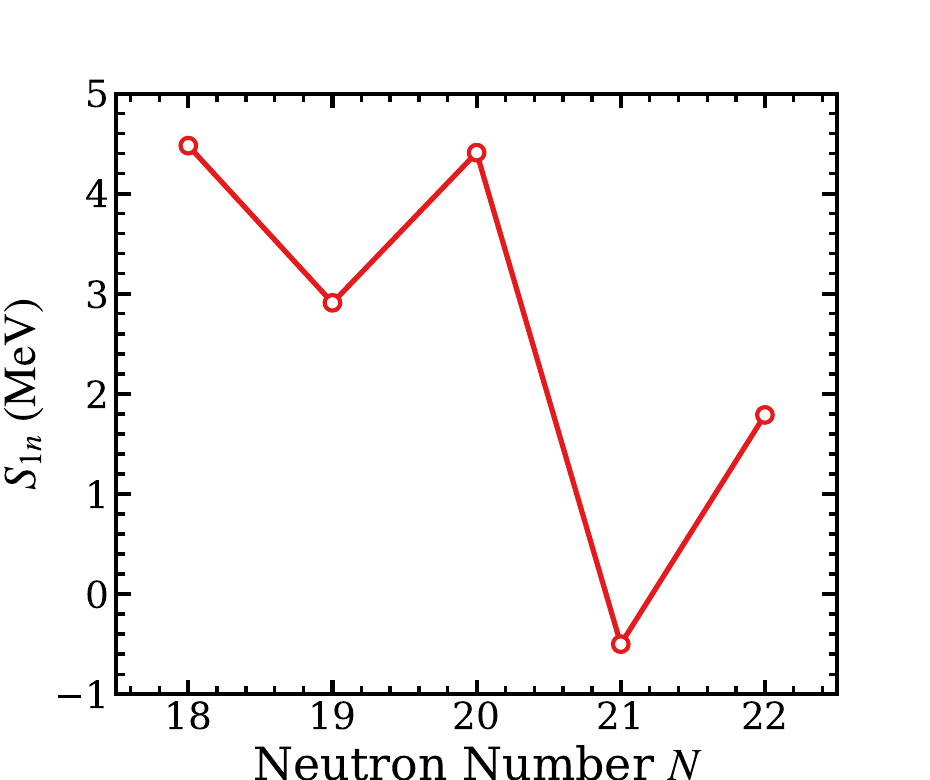}
\caption{(Color online) One-neutron separation energies $S_{1n}$ of $^{28\text{--}32}$Ne calculated within the DRHBc framework
as functions of neutron number $N$. A pronounced drop of $S_{1n}$ is observed at $N=21$
($^{31}$Ne), indicating extremely weak binding of the valence neutron. The odd--even staggering pattern is clearly visible along the isotopic chain.}
\label{fig:sn_drhbc}
\end{figure}

The microscopic densities analyzed in this work are obtained from the deformed relativistic Hartree--Bogoliubov theory in continuum (DRHBc), originally formulated in Refs.~\cite{Zhou2010,Li2012} and recently extended to large-scale applications in Ref.~\cite{DRHBc2024}. This framework treats pairing correlations, deformation, and continuum coupling in a fully self--consistent manner. The DRHBc approach has been successfully applied to the description of halo phenomena in a wide range of light and medium--\blu{ and heavy--mass} nuclei \cite{Zhou2010,Li2012,DRHBc2024,Wang2024EPJA,An2024PLB}, \red{providing a reliable starting point for the present study.} \red{Recent DRHBc+Glauber studies of one-neutron halo systems in the C and Mg chains have also shown that microscopic densities can be linked consistently to reaction observables within a unified framework \cite{Wang2024EPJA,An2024PLB}.} The one-neutron separation energies calculated within the
DRHBc framework are shown in Fig.~\ref{fig:sn_drhbc}.
A clear odd--even staggering pattern is observed.
In particular, $^{31}$Ne ($N=21$) exhibits a drastic reduction
of $S_{1n}$, approaching the weak-binding regime,
whereas the neighboring isotopes remain more strongly bound \cite{DRHBc2024}. 

For the Ne isotopes considered here, the DRHBc calculations predict a coexistence of spherical and deformed shapes. Specifically, $^{28}$Ne, $^{30}$Ne, and $^{32}$Ne are found to be spherical, while $^{29}$Ne and $^{31}$Ne exhibit axial deformation as summarized in Table~\ref{tab:drhbc_basic}. This mixture of shapes within a narrow isotopic chain makes Ne isotopes particularly suitable for investigating the interplay between deformation and halo formation.
For the deformed nuclei $^{29}$Ne and $^{31}$Ne, the matter density distributions are expanded in terms of Legendre polynomials with multipolarities $\lambda = 0, 2, 4, 6$ as
\begin{equation}
\rho(r,\theta) = \sum_{\lambda = 0,2,4,6,\dots} \rho_{\lambda}(r) P_{\lambda}(\cos\theta).
\label{eq:rho_legendre}
\end{equation}
Using this expansion, we analyze the matter density distributions \blu{in the following sections.}

\begin{table}[t]
\caption{Basic structural parameters of $^{28\text{--}32}$Ne obtained
from the DRHBc calculations. The parameter $R_{\mathrm{rms}}$ denotes
the root--mean--square radius, and $\beta_2$ denotes
the quadrupole deformation parameter \blu{from Ref. \cite{DRHBc2024}.}}
\label{tab:drhbc_basic}
\centering
\begin{tabular}{cccccc}
\hline\hline
Nucleus
& $^{28}$Ne & $^{29}$Ne & $^{30}$Ne & $^{31}$Ne & $^{32}$Ne \\
\hline
$R_{\mathrm{rms}}$ (fm)
& 3.167 & 3.234 & 3.288 & 3.499 & 3.430 \\
$\beta_2$
& 0.000 & 0.063 & 0.000 & 0.177 & 0.000 \\
\hline\hline
\end{tabular}
\end{table}

\subsection{Spatial features of the density distributions}    

The matter density distributions in Table~\ref{tab:density_xz} in the intrinsic frame reveal clear qualitative differences among the Ne isotopes. For $^{28}$Ne and $^{30}$Ne, the density decreases relatively rapidly beyond the nuclear surface, and the low--density tail remains limited. In contrast, $^{31}$Ne exhibits a pronounced spatial extension, with the density persisting to significantly larger radii. This extension is anisotropic and reflects the combined effects of weak binding and deformation.

The case of $^{29}$Ne is more subtle. Although this nucleus is deformed, its density tail does not show a dramatic enhancement compared to its even--even neighbors. For $^{32}$Ne, the overall nuclear size increases, but the distinction between the bulk and the tail regions becomes less clear, making it difficult to assess whether the observed extension originates from a genuine halo or from a relatively thick neutron skin. \red{This ambiguity underscores the difficulty of distinguishing a genuine halo from a thick neutron skin on the basis of global geometric measures alone.}

\begin{table}[htbp]
    \caption{(Color online) Matter density distributions of $^{28\text{--}32}$Ne in the $x$--$z$ plane obtained from the DRHBc calculations. Root--mean--square radii $R_{\mathrm{rms}}$ and tail ratios
$I_2/(I_1+I_2)$ \blu{in Eq. (\ref{eq:I1_I2})} for $^{28\text{--}32}$Ne obtained from the DRHBc
matter density distributions. The separation radius in Eq.~(\ref{eq:I1_I2}) is taken as the DRHBc rms radius $R_{\mathrm{rms}}$.}
    \label{tab:density_xz}
    \centering
    \renewcommand{\arraystretch}{1.2}
    \setlength{\tabcolsep}{10pt}

    \begin{ruledtabular}
    \begin{tabular}{c c c c c c}
        Nucleus & $^{28}\mathrm{Ne}$ & $^{29}\mathrm{Ne}$ & $^{30}\mathrm{Ne}$ & $^{31}\mathrm{Ne}$ & $^{32}\mathrm{Ne}$ \\
        $R_{\mathrm{rms}}\ \mathrm{(fm)}$ & 3.167 & 3.234 & 3.288 & 3.499 & 3.430 \\
        $I_2 /(I_1+I_2)$ & 0.4050 & 0.4029 & 0.4039 & 0.3546 & 0.3925 \\
        \raisebox{1.15cm}{Images}
        & \includegraphics[width=0.13\textwidth]{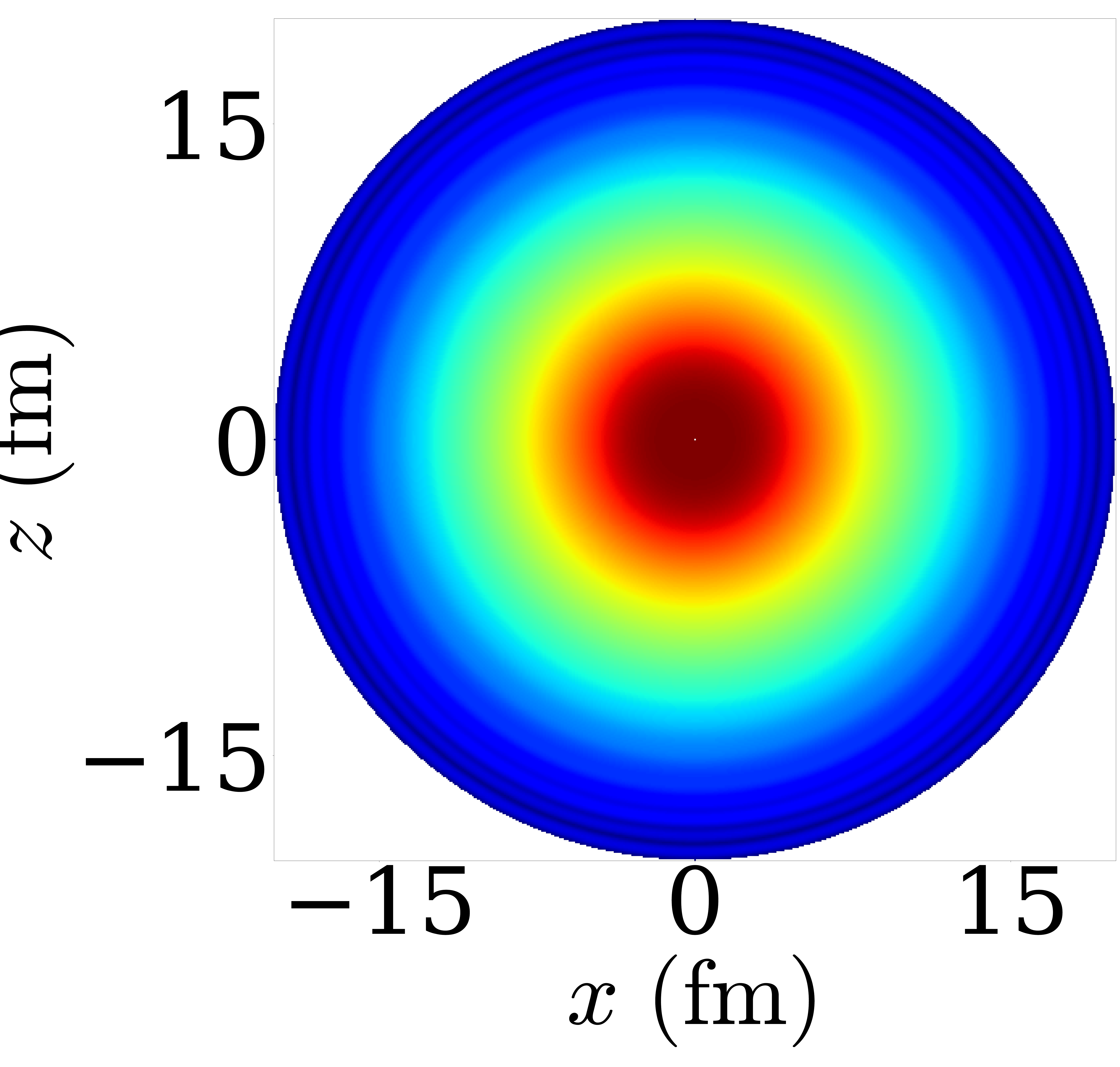}
        & \includegraphics[width=0.13\textwidth]{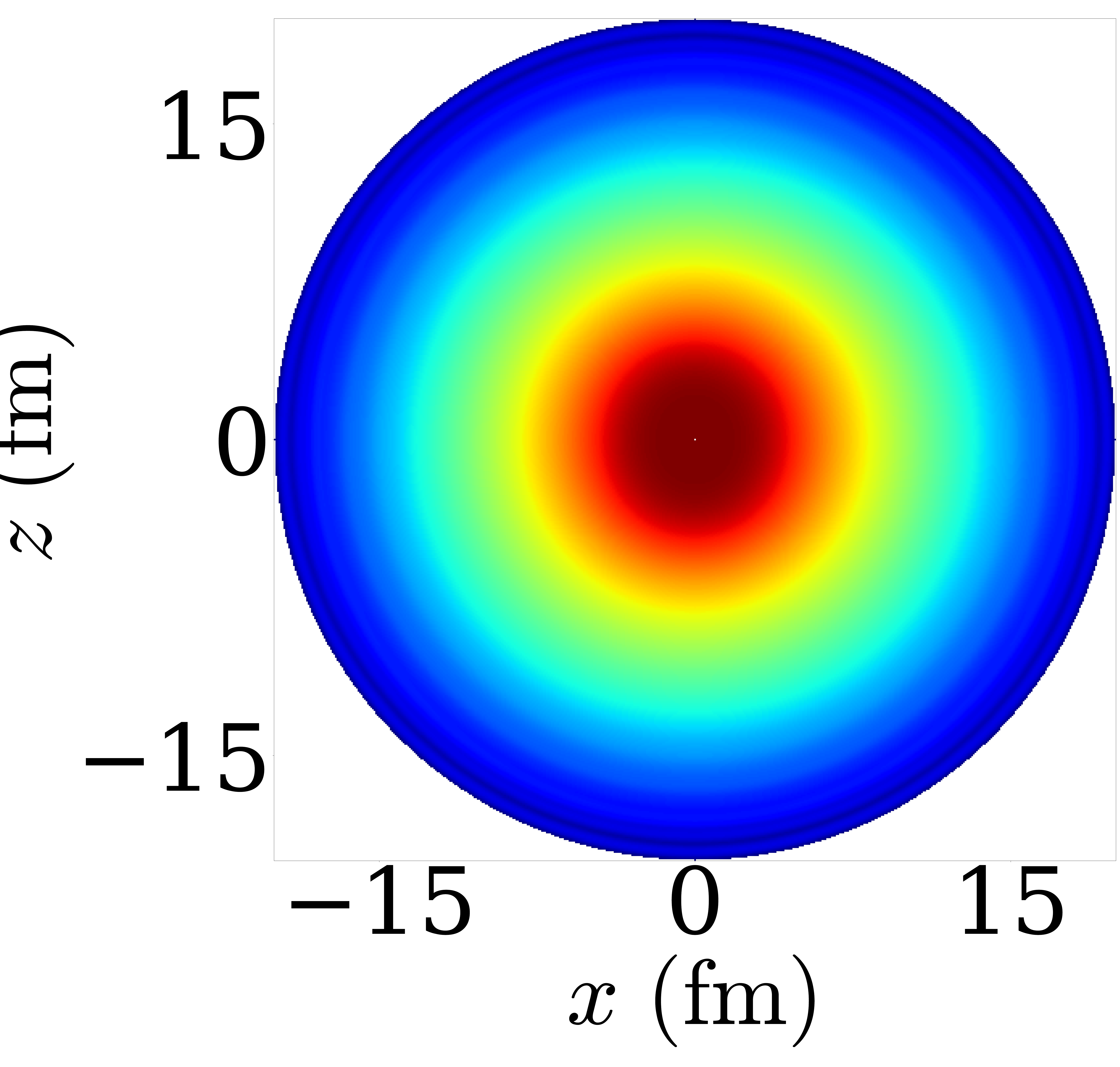}
        & \includegraphics[width=0.13\textwidth]{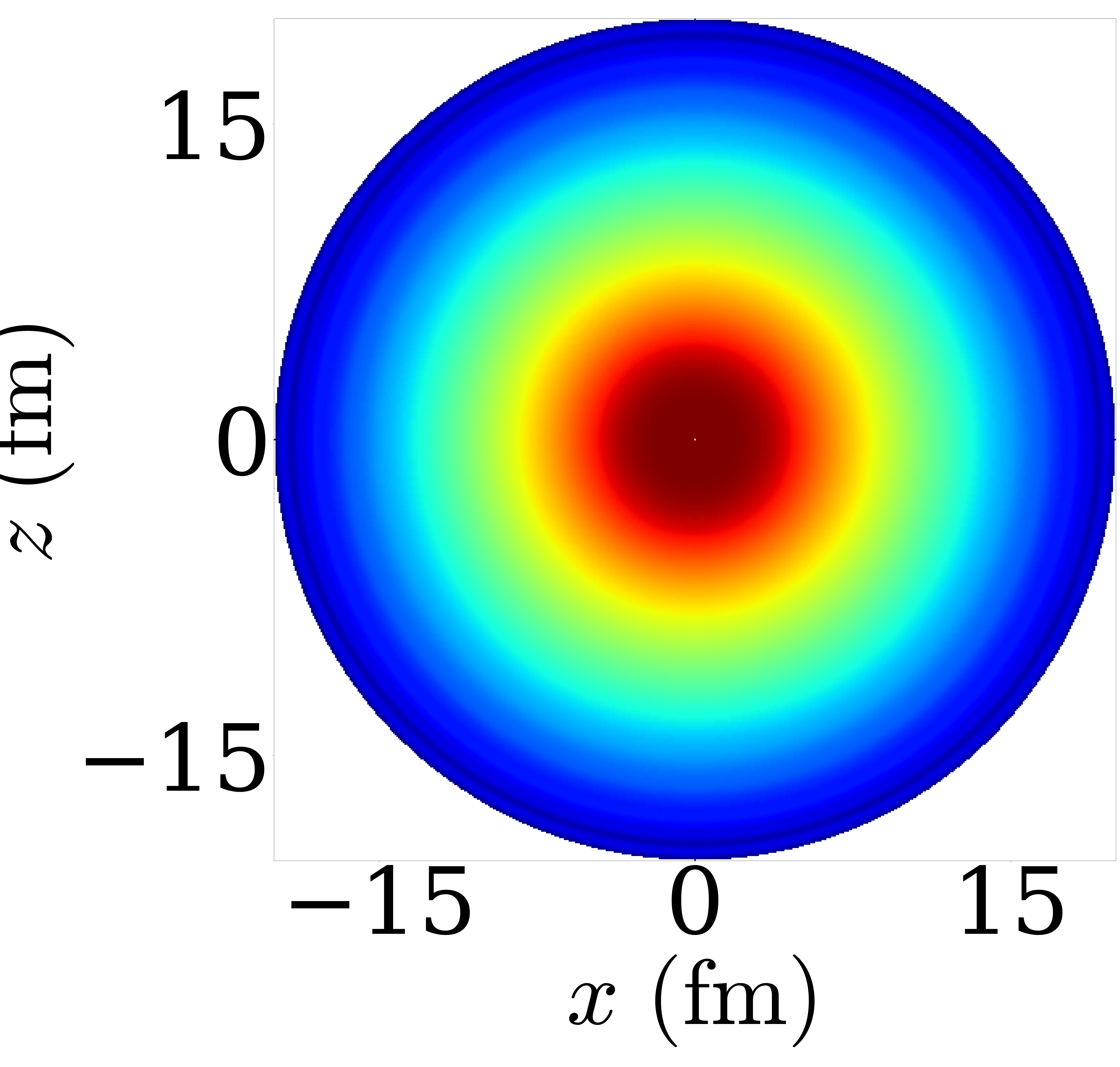}
        & \includegraphics[width=0.13\textwidth]{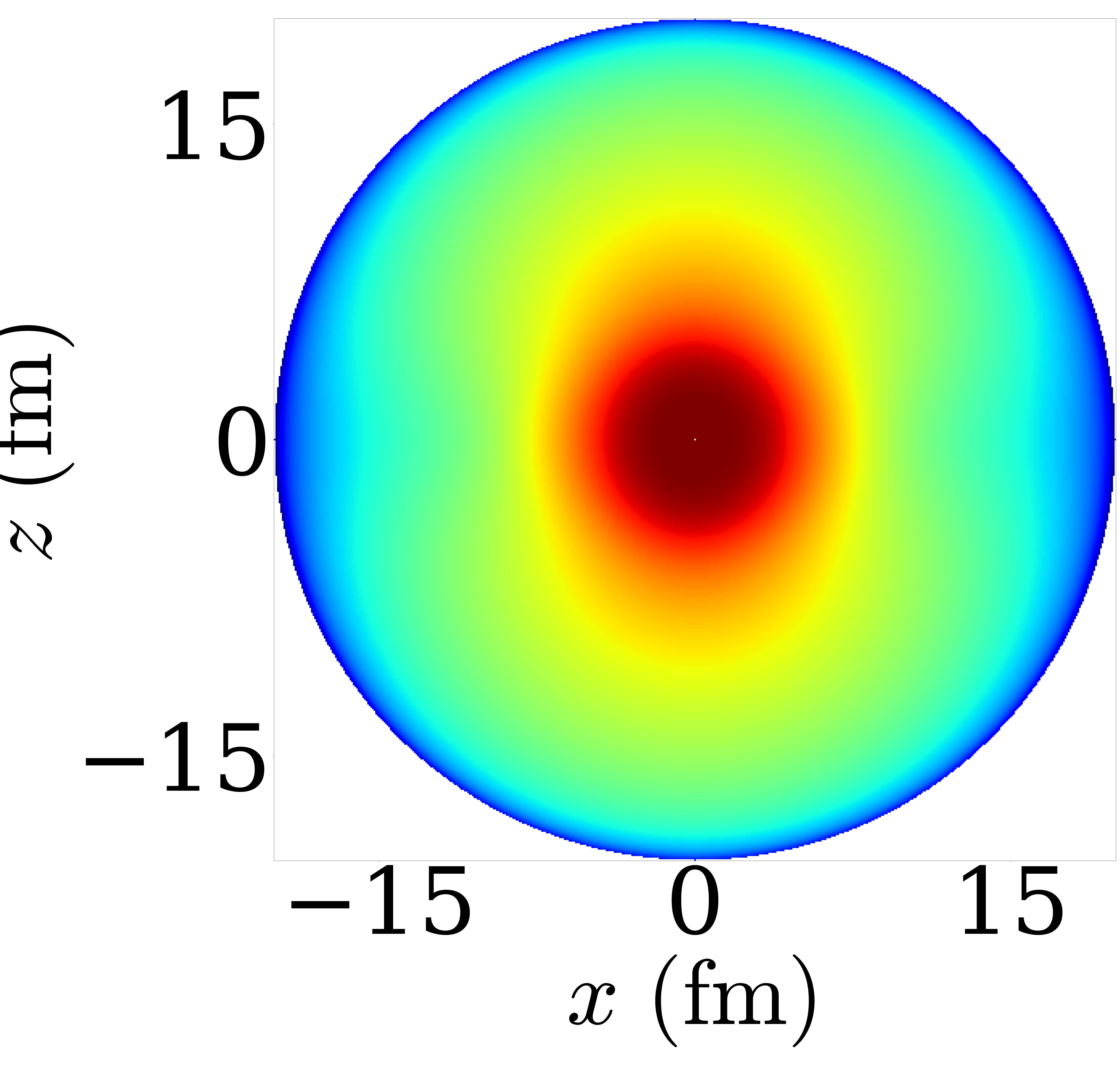}
        & \includegraphics[width=0.13\textwidth]{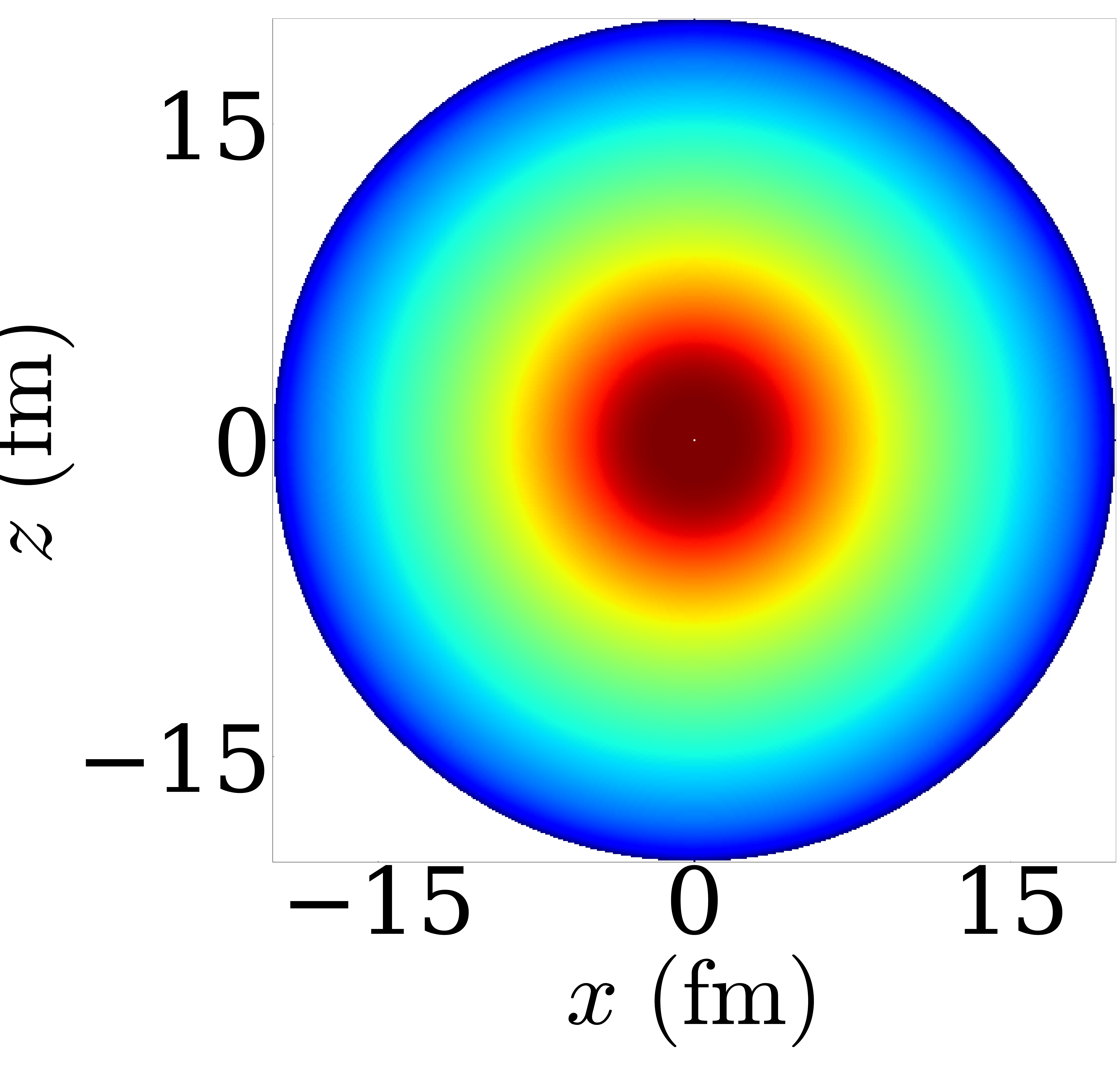}
         \\
         \\[-6pt]
         \multicolumn{6}{c}{\includegraphics[width=0.25\textwidth]{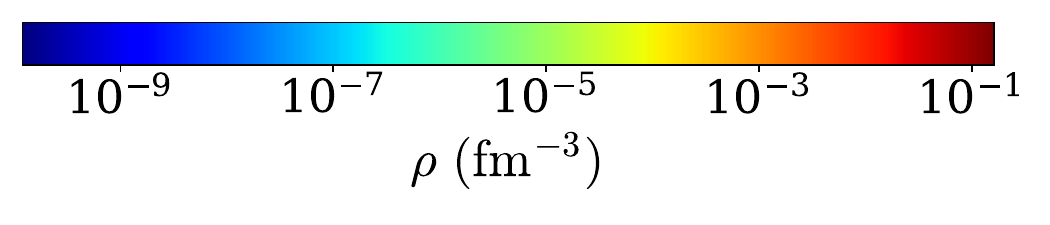}}
    \end{tabular}
    \end{ruledtabular}
\end{table}

\subsection{Quantifying tail contributions using $R_{\mathrm{rms}}$}

To quantify the contribution of the outer region of the density distribution, we first adopt the rms radius $R_{\mathrm{rms}}$ as a separation scale and define the integrals
\begin{equation}
I_1 = \int_0^{R_{\mathrm{rms}}} \rho(r) d^3r, \qquad
I_2 = \int_{R_{\mathrm{rms}}}^{20,\mathrm{fm}} \rho(r) d^3r,
\label{eq:I1_I2}
\end{equation}
where $\rho(r)$ denotes the matter density. The ratio $I_2/(I_1+I_2)$ represents the fractional contribution of the tail region.

An important observation is that, for halo nuclei, the rms radius itself is already enlarged by the extended tail. As a result, the ratio $I_2/(I_1+I_2)$ does not necessarily increase and may even decrease for nuclei with pronounced halos. This behavior is indeed found for $^{31}$Ne, which exhibits the largest $R_{\mathrm{rms}}$ among the isotopes considered, yet does not show a correspondingly large tail fraction when $R_{\mathrm{rms}}$ is used as the separation radius. This limitation indicates that $R_{\mathrm{rms}}$ is not an optimal scale for isolating halo contributions.

\subsection{Neutron density analysis and the $R_{\rho_0/2}$ criterion}

\begin{table}[t] 
\centering
\caption{Results of the tail-contribution analysis for the neutron density distributions using $R^n_{\rho_0/2}$ as the separation radius.}
\label{tab:Rrho_half_neutron}
\begin{ruledtabular}
\begin{tabular}{lccccc}
Nuclei & $^{28}$Ne & $^{29}$Ne & $^{30}$Ne & $^{31}$Ne & $^{32}$Ne \\
\hline
$R_{\rho_0/2,n}$ & 4.216 & 4.289 & 4.362 & 4.385 & 4.480 \\
$I_2/(I_1+I_2)$  & 0.161 & 0.164 & 0.163 & 0.191 & 0.180 \\
\end{tabular}
\end{ruledtabular}
\end{table}

To overcome the ambiguity associated with $R_{\mathrm{rms}}$, we introduce an alternative separation radius based on the neutron density distribution. Focusing on the monopole component of the neutron density $\rho^n(r)$, we define the central density \blu{$\rho^n_{0} = \rho^n (0)$ and determine the radius $R^n_{\rho_0/2}$ by the condition $\rho^n(R^n_{\rho_0/2}) = \rho^n_{0}/2$.} Using this radius as the boundary between the inner and outer regions, the integrals $I_1$ and $I_2$ are re--evaluated for the neutron density distribution.

The resulting tail contributions are summarized in Table~III. For $^{28}$Ne, $^{29}$Ne, and $^{30}$Ne, the ratio $I_2/(I_1+I_2)$ remains nearly constant at about 0.16, indicating no significant enhancement of the neutron tail. In contrast, $^{31}$Ne exhibits a clear increase of the tail contribution to $I_2/(I_1+I_2)=0.191$, which is distinctly larger than the values of its neighboring isotopes. This enhancement reflects the presence of a spatially extended neutron density beyond $R_{\rho_0/2}$.

For $^{32}$Ne, the neutron tail contribution also increases compared to the lighter isotopes, reaching $I_2/(I_1+I_2)=0.180$. However, this enhancement is weaker than that observed for $^{31}$Ne and does not constitute an unambiguous signature of a halo structure. At the level of microscopic neutron densities, $^{31}$Ne therefore stands out as the most prominent case with an anomalously extended tail, while the nature of the spatial extension in $^{32}$Ne remains uncertain.

\subsection{Global halo indicators and comparison with the diffuseness systematics}
\label{subsec:global_indicators}

For completeness, we also examine two global halo indicators proposed in studies of medium-mass halos, namely the average number of neutrons in the spatially decorrelated region $N_{\mathrm{halo}}$~\cite{Rotival2009} and the one-neutron halo scale $S_{\mathrm{halo\text{-}1n}}$~\cite{Zhang2023}.
These indicators do not require an explicit \emph{a priori} separation of the nucleus
into a core and a halo, and are therefore useful for screening halo candidates. Following the definition introduced in \blu{Ref.~\cite{Zhang2023},} the one-neutron halo scale is
given by
\begin{equation}
S_{\mathrm{halo\text{-}1n}}
=
\frac{\Delta \tilde{R}_n}{\Delta R^{\mathrm{emp}}_n}
=
\frac{\tilde{R}_n(N)-\tilde{R}_n(N-1)}
     {R^{\mathrm{emp}}_n(N)-R^{\mathrm{emp}}_n(N-1)} ,
\label{eq:Shalo}
\end{equation}
where $N$ denotes the neutron number, and
$R^{\mathrm{emp}}_n=r_0 N^{1/3}$ is the empirical neutron radius with $r_0=1.22$ fm.
To reduce deformation-induced ambiguities, we adopt the spherical-reduced neutron radius
\begin{equation}
\tilde{R}_n = \left(1+\frac{5\beta_{2n}^2}{4\pi}\right)^{-1/2} R_n ,
\label{eq:Rtilde}
\end{equation}
where $R_n$ and $\beta_{2n}$ are the DRHBc neutron rms radius and neutron quadrupole deformation,
respectively. The second global indicator~\cite{Rotival2009} is defined as the average number of neutrons contained in
the decorrelated outer region,
\begin{equation}
N_{\mathrm{halo}}
=
4\pi \int_{r_0}^{\infty} \rho^n_{0}(r)\, r^2\, dr ,
\label{eq:Nhalo}
\end{equation}
where $\rho^n_{0}(r)$ is the angle-averaged neutron density.
The lower limit $r_0$ determines the onset of the halo-like region in the sense of Ref.~\cite{Rotival2009}; in practical implementations, it may be chosen from a density-curvature criterion in the tail region.

\begin{figure}[htbp]
    \centering
    
    \begin{subfigure}{0.48\textwidth}
        \centering
        \includegraphics[width=\linewidth]{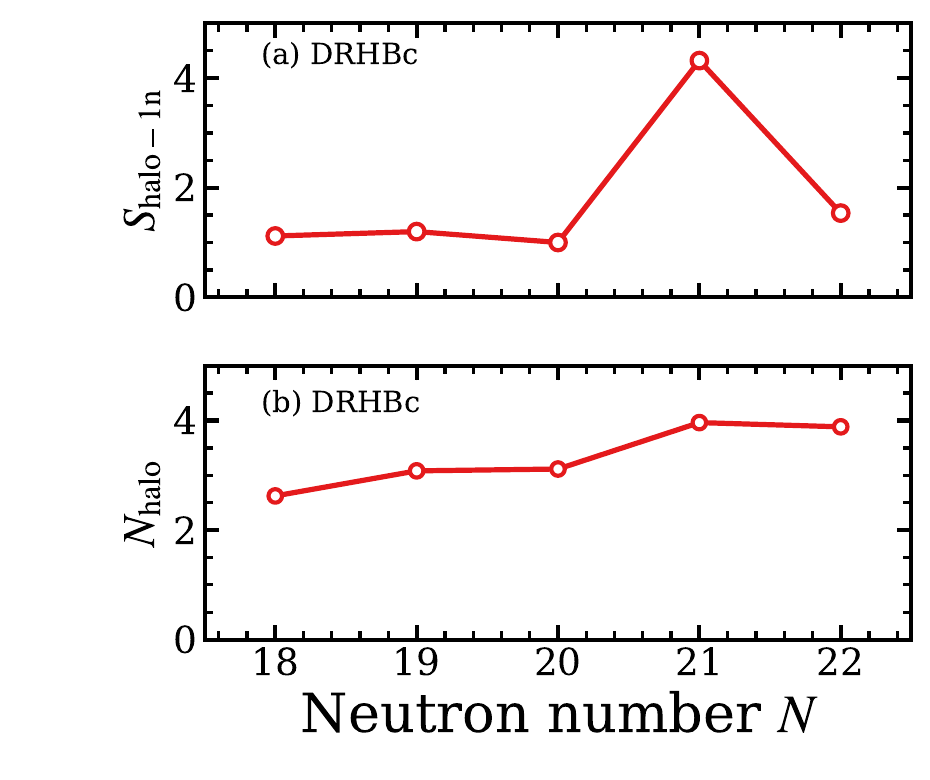}
        
        \label{fig:fig_Shalo_Nhalo}
    \end{subfigure}
    \hfill
    \begin{subfigure}{0.48\textwidth}
        \centering
        \includegraphics[width=\linewidth]{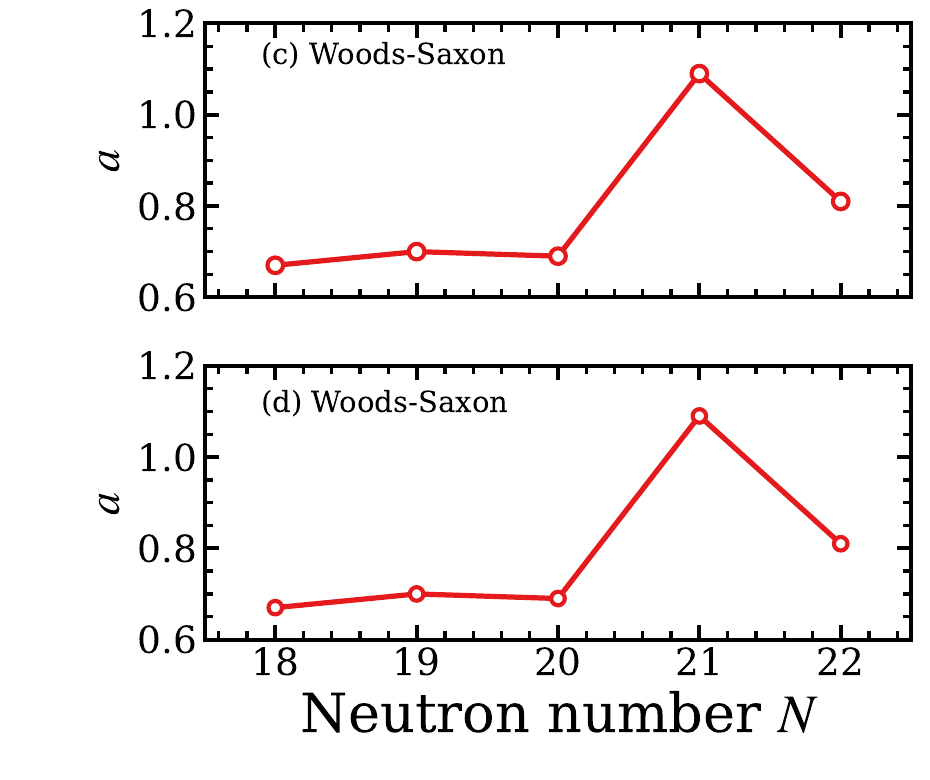}
      
        \label{fig:a_halo}
    \end{subfigure}

    \caption{(Color online) Global halo indicators and diffuseness systematics for
  $^{28\text{--}32}$Ne as a function of neutron number $N$.
  Panels (a) and (b) show the one-neutron halo scale $S_{\mathrm{halo\text{-}1n}}$
  and the decorrelated neutron number $N_{\mathrm{halo}}$, respectively, evaluated
  from the DRHBc densities following Eqs.~(\ref{eq:Shalo}) and (\ref{eq:Nhalo}).
  Panels (c) and (d) display the Woods--Saxon diffuseness parameter '$a$' extracted
  from the logarithmic-scale WS fitting for the two normalization choices $\rho_0 = \rho(r=0)$ and $\rho_0 = \rho_{\max}$
  (see also Sec.~III and \blu{Table~\ref{tab:ws_fit_r0}}).
  A pronounced anomaly is observed at $N=21$ ($^{31}$Ne), where both
  $S_{\mathrm{halo\text{-}1n}}$ and $a$ exhibit their largest values, while $^{32}$Ne
  shows a moderate enhancement consistent with an intermediate, extended structure.}
    \label{fig:halo_comparison}
\end{figure}

Figure~\ref{fig:halo_comparison} summarizes the global indicators
$S_{\mathrm{halo\text{-}1n}}$ and $N_{\mathrm{halo}}$ obtained for $^{28\text{--}32}$Ne,
together with the Woods--Saxon diffuseness parameter $a$ extracted in Sec.~III (shown
here for comparison).
A clear anomaly is observed at $N=21$ ($^{31}$Ne), where $S_{\mathrm{halo\text{-}1n}}$
exhibits a pronounced enhancement and the diffuseness reaches its maximum value.
The quantity $N_{\mathrm{halo}}$ shows an overall increasing trend toward the neutron-rich
side and is also enhanced around $^{31,32}$Ne.
These observations qualitatively support the identification of $^{31}$Ne as the most
prominent halo candidate within the present isotopic chain, while $^{32}$Ne displays an
intermediate behavior.
This motivates the more quantitative, locally defined criterion based on the surface
diffuseness developed in the next section.

\section{Phenomenological criterion based on surface diffuseness}


The analysis of microscopic density distributions in Sec.~II
demonstrates that halo structures manifest themselves primarily
through anomalously extended low--density tails.
However, global observables such as the rms radius or tail
fractions defined with respect to $R_{\mathrm{rms}}$ are not always
sufficient to unambiguously characterize halo nuclei, especially
in the presence of deformation.
This limitation motivates the introduction of a locally defined
and quantitative measure that is directly sensitive to the radial
falloff of the density distribution.

In this context, the surface diffuseness parameter of a
Woods--Saxon (WS) form provides a natural candidate.
The diffuseness characterizes how gradually the
density decreases at the nuclear surface.
For halo nuclei, the weak binding of valence nucleons leads to a
slowly decaying density profile, which is expected to be reflected
in an unusually large diffuseness parameter. \red{Physically, the diffuseness parameter controls the radial slope in the surface region. For weakly bound systems, reduced centrifugal-barrier effects and continuum coupling lead to a slower falloff, which manifests itself as enhanced diffuseness.} \red{A related sensitivity of phenomenological radius/diffuseness parameters to halo-induced long-range structure has also been noted in optical-model analyses of light halo reactions \cite{Kim2018,Heo2020}.}

\subsection{Deformed Woods--Saxon fitting procedure}

To quantify the surface diffuseness, we fit the DRHBc density
distributions with a deformed Woods--Saxon form,
\begin{equation}
\rho(r,\theta) =
\frac{\rho_0}
{1+\exp\!\left[\dfrac{r-R(\theta)}{a}\right]},
\label{eq:ws_density}
\end{equation}
where $a$ denotes the surface diffuseness and
\begin{equation}
R(\theta) = R_0\left[1+\beta_2 Y_{20}(\theta)\right]
\label{eq:ws_radius}
\end{equation}
accounts for axial quadrupole deformation.
In the present analysis, the parameters, \blu{$\beta_2$ in Table~\ref{tab:drhbc_basic} and $\rho_0$ in Table~\ref{tab:ws_fit_r0},} are fixed to the values obtained from the DRHBc calculations, while the diffuseness parameter $a$ and the radius parameter $R_0$ are varied.

Since the halo signature is encoded mainly in the dilute tail
region of the density distribution, the fitting procedure is
performed in logarithmic scale using the mean squared logarithmic
error.
This choice ensures that the low--density region is weighted
appropriately and that the asymptotic behavior of the density is
faithfully reproduced. The normalization parameter $\rho_0$ is taken either as the maximum
density $\rho_{\max}$ or as the \blu{central} density at $r=0$ of the DRHBc distribution. The optimal values of the WS parameters are determined by
minimizing the fitting error under the normalization constraint
\begin{equation}
A_{\mathrm{WS}} = \int \rho(r,\theta)\, d^3 r .
\end{equation}

\begin{figure}[htbp]
    \centering
    
    \begin{subfigure}{0.48\textwidth}
        \centering
        \includegraphics[width=\linewidth]{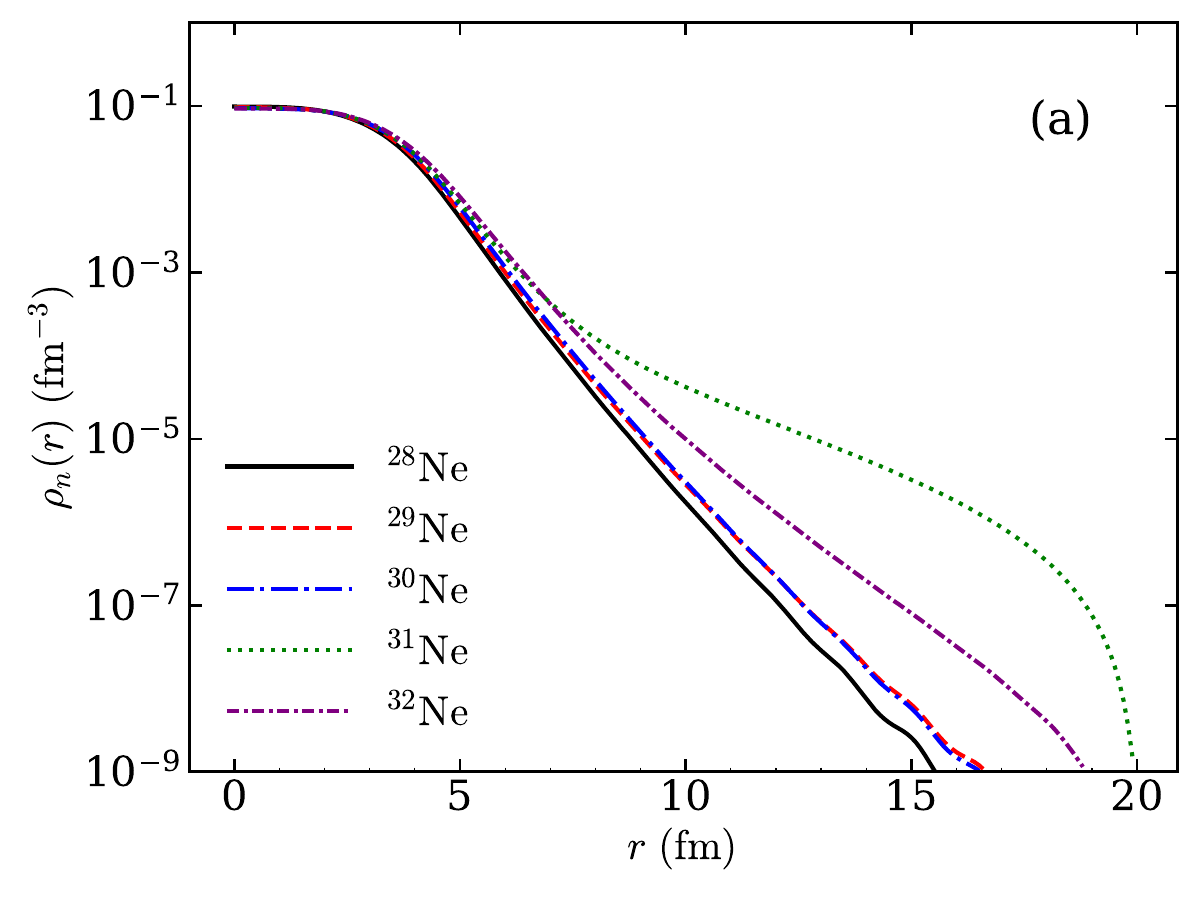}
        
        \label{fig:DRHBc_neutron_distribution}
    \end{subfigure}
    \hfill
    \begin{subfigure}{0.48\textwidth}
        \centering
        \includegraphics[width=\linewidth]{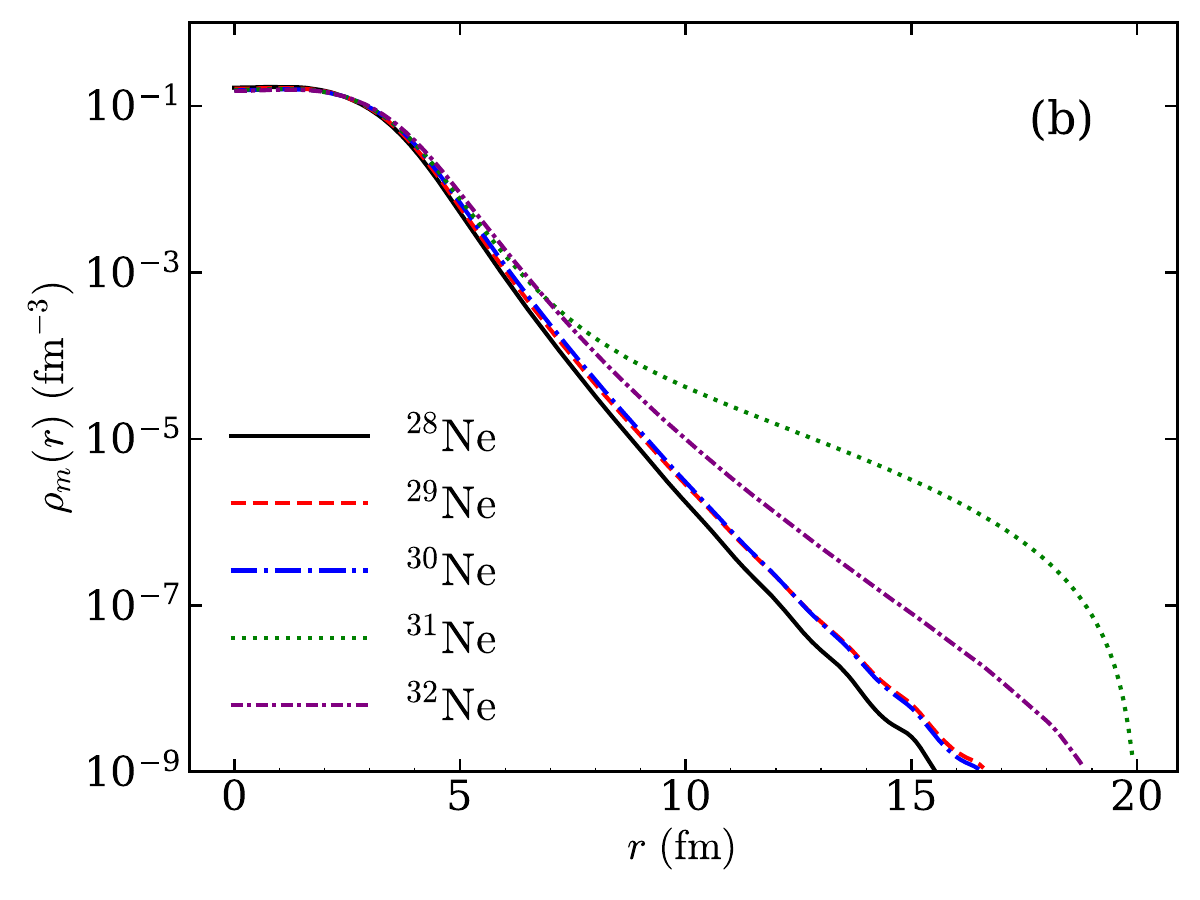}
      
        \label{fig:DHRBc_matter_distribution}
    \end{subfigure}

    \caption{(Color online) Angle-averaged density distributions of
  $^{28\text{--}32}$Ne obtained from the DRHBc calculations in logarithmic scale.
  The left panel shows the neutron density $\rho_n(r)$ and the right panel shows
  the matter density $\rho_m(r)$ as functions of the radius $r$.}
  \label{fig:drhbc_density_log}
\end{figure}

Before introducing the phenomenological criterion,
it is instructive to examine the radial behavior of the
microscopic density distributions in more detail.
Figure~\ref{fig:drhbc_density_log} displays the angle-averaged
neutron and matter density distributions of
$^{28\text{--}32}$Ne in logarithmic scale.
While the densities of $^{28}$Ne, $^{29}$Ne, and $^{30}$Ne
decrease approximately in parallel beyond the surface region,
$^{31}$Ne exhibits a markedly slower falloff at large radii.
The tail of $^{32}$Ne is also moderately enhanced,
but remains clearly below that of $^{31}$Ne.
This behavior suggests that the essential difference among
these isotopes lies in the radial steepness of the surface region.

\begin{figure}[htbp]
    \captionsetup[subfigure]{labelformat=empty}
    \centering
    
    \begin{subfigure}{0.30\textwidth}
        \centering
        \includegraphics[width=\linewidth]{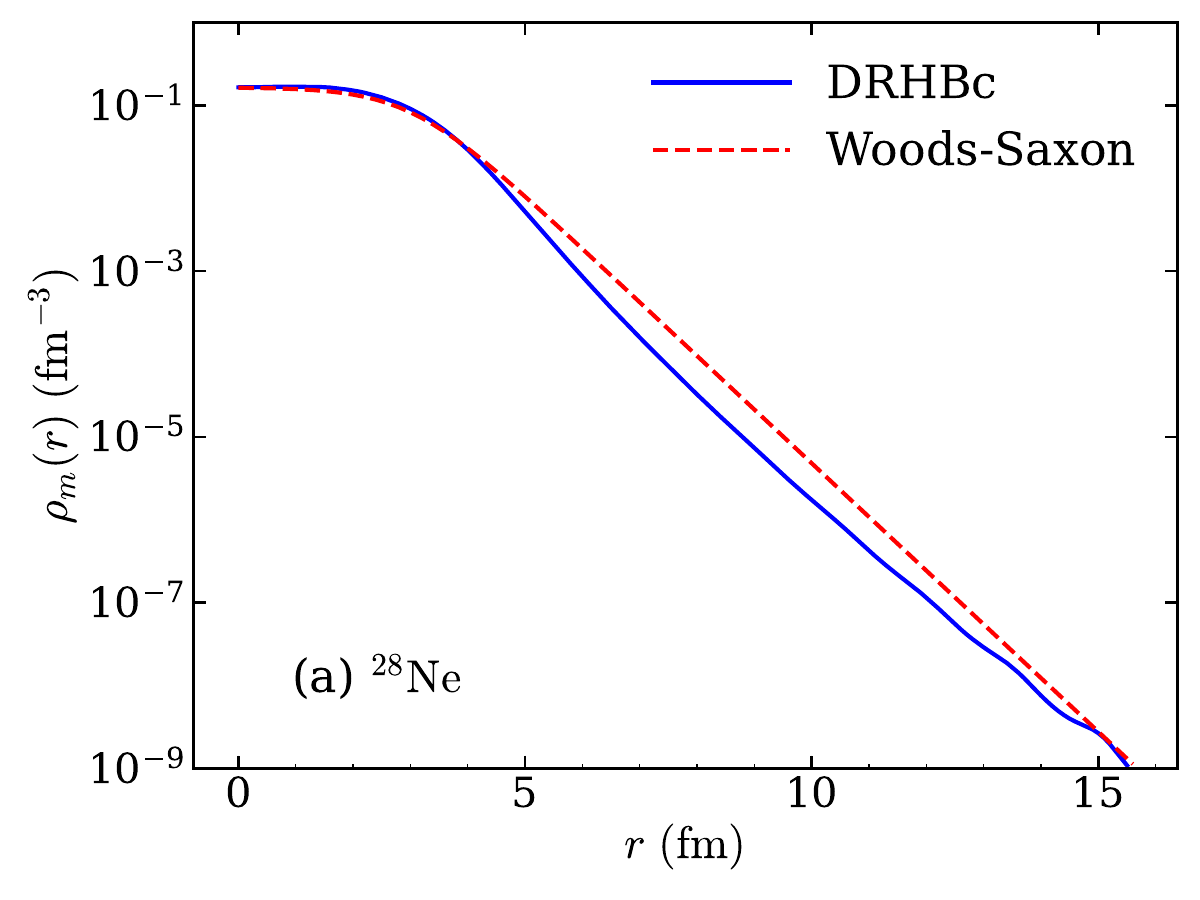}
    \end{subfigure}
    \hfill
    \begin{subfigure}{0.30\textwidth}
        \centering
        \includegraphics[width=\linewidth]{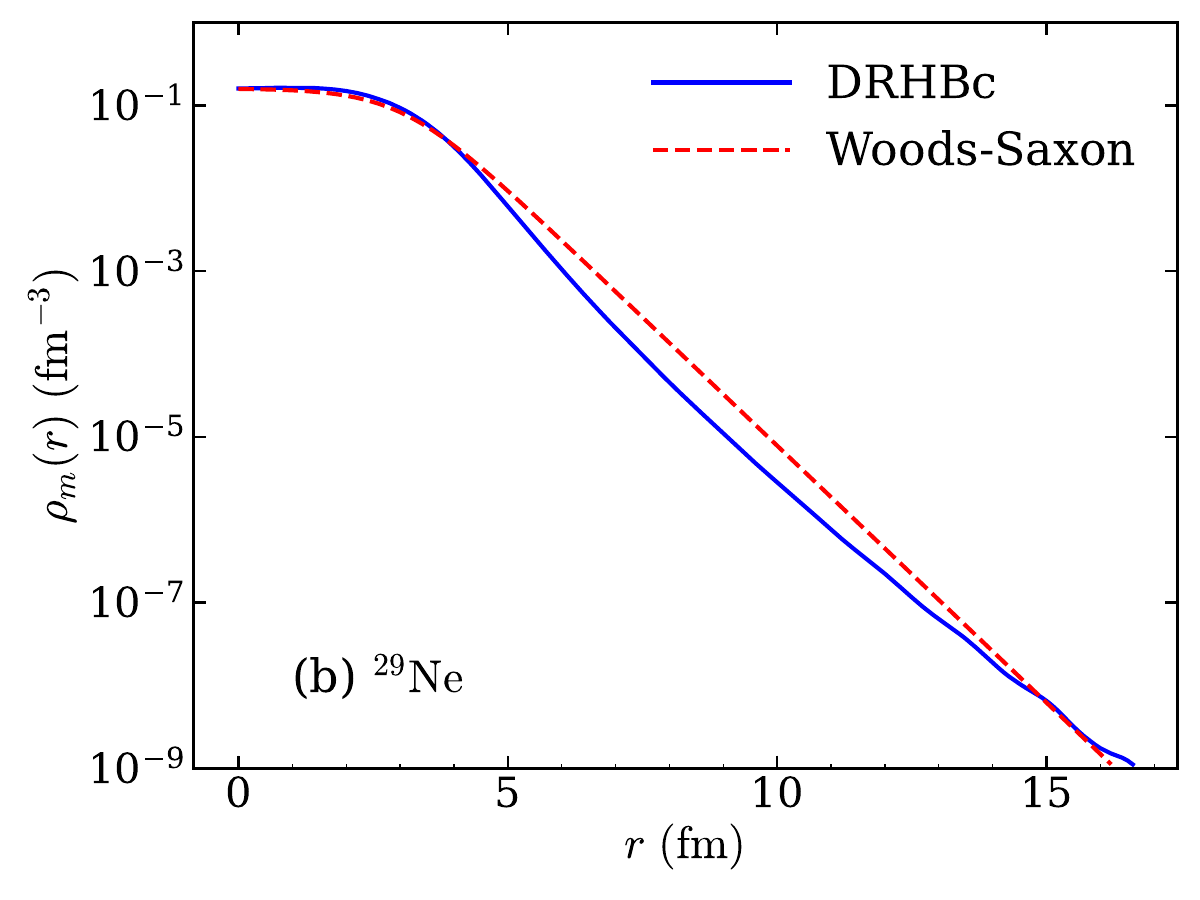}
    \end{subfigure}
    \hfill
    \begin{subfigure}{0.30\textwidth}
        \centering
        \includegraphics[width=\linewidth]{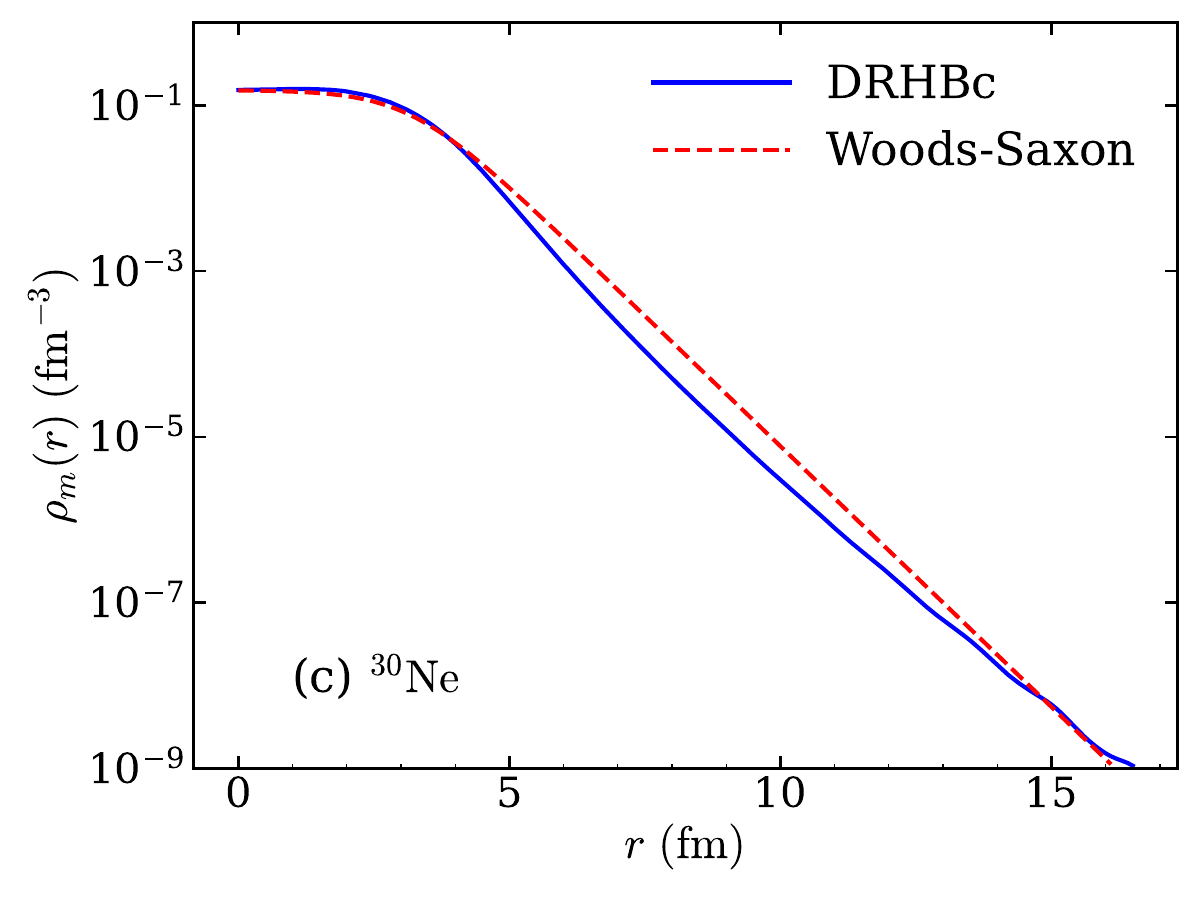}
    \end{subfigure}

    \vspace{0.5cm}

    \makebox[\textwidth][c]{%
        \begin{subfigure}{0.30\textwidth}
            \centering
            \includegraphics[width=\linewidth]{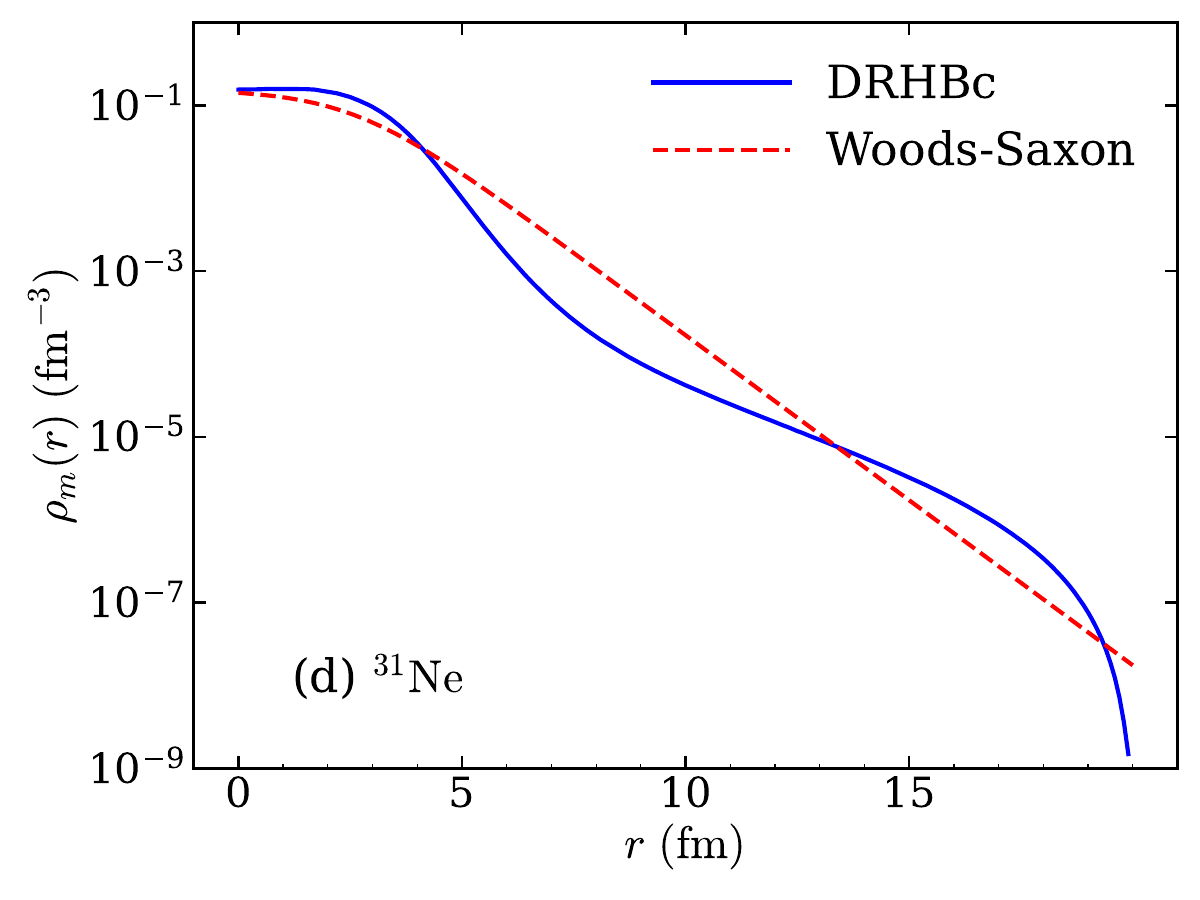}
        \end{subfigure}
        \hspace{0.05\textwidth}
        \begin{subfigure}{0.30\textwidth}
            \centering
            \includegraphics[width=\linewidth]{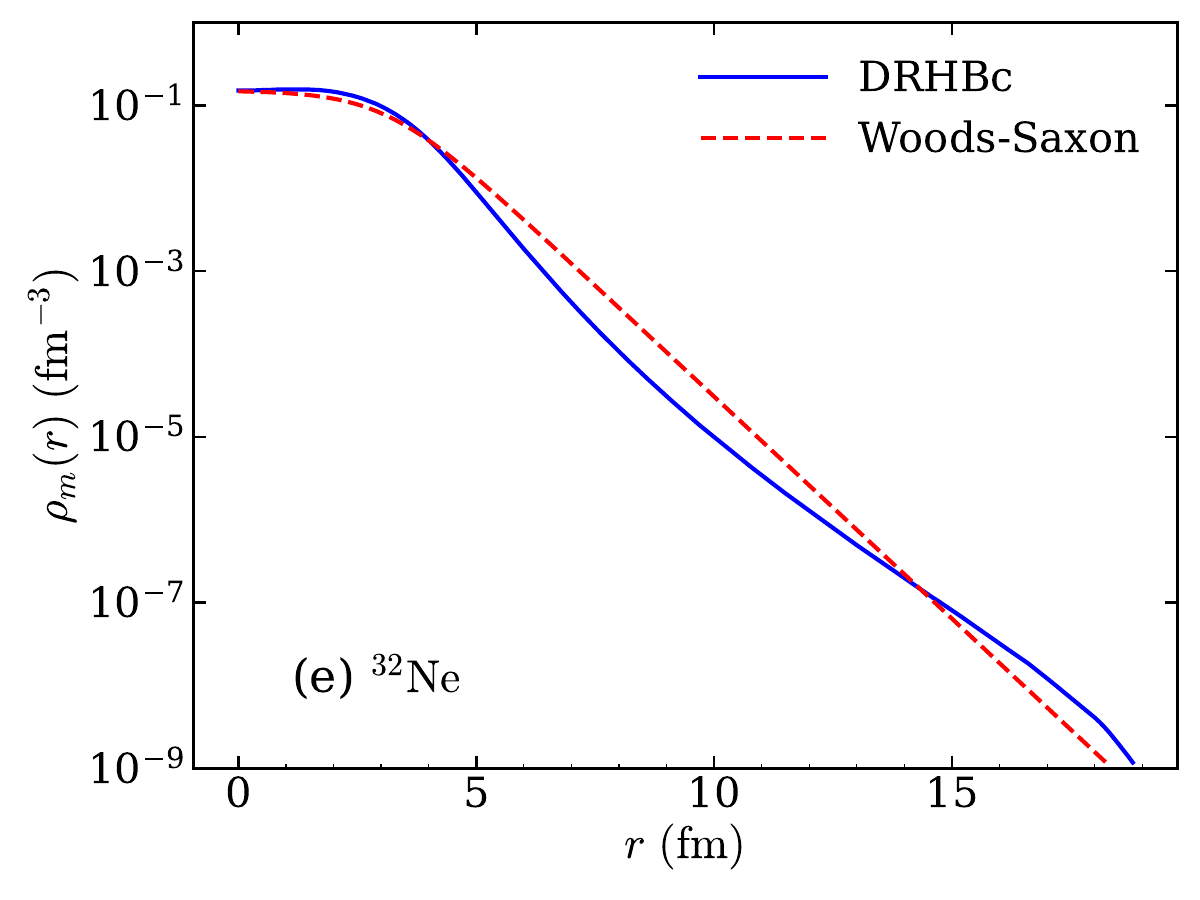}
        \end{subfigure}
    }
    \caption{(Color online) Comparison between the DRHBc matter density distributions
  and the deformed WS fits for $^{28\text{--}32}$Ne in logarithmic scale.
  In each panel, the DRHBc density (\blu{blue}) is compared with the WS parametrization (red)
  optimized using the logarithmic-scale fitting procedure.}
  \label{fig:ws_fit_panels}

\end{figure}

The quality of the WS fitting is illustrated in
Fig.~\ref{fig:ws_fit_panels},
where the DRHBc density distributions are compared with the
optimized WS parametrizations in logarithmic scale.
For $^{28\text{--}30}$Ne, the WS curves closely follow the
microscopic densities over the entire radial range.
In the case of $^{31}$Ne, however, the extended tail of the
microscopic density requires a substantially larger diffuseness
parameter to reproduce the slow radial decay. \red{A pronounced enhancement of the low-density tail is observed for $^{31}$Ne, while $^{32}$Ne exhibits only a moderate extension compared with the lighter isotopes.}
This behavior directly reflects the weak binding of the
valence neutron and the resulting long-range density profile.
The deviation between isotopes is therefore encoded
not in the central density, but in the asymptotic slope.

\subsection{Diffuseness systematics in $^{28\text{--}32}$Ne}

The resulting WS fitting parameters are summarized in \blu{Tables~\ref{tab:ws_fit_r0}}. For $^{28}$Ne, $^{29}$Ne, and $^{30}$Ne, the extracted diffuseness parameters remain within a narrow range, $a \simeq 0.67\text{--}0.70$~fm, which is typical of ordinary medium--mass nuclei. \red{For $^{31}$Ne, the diffuseness rises sharply to $a \simeq 1.09$~fm for both normalization choices, making the large diffuseness the clearest phenomenological anomaly in the isotopic chain.} For $^{32}$Ne, the diffuseness increases to about $a \simeq 0.81$~fm, larger than that of the lighter isotopes but still significantly smaller than that of $^{31}$Ne.

\begin{figure*}[htbp]
    \centering

    \begin{subfigure}[b]{0.17\textwidth}
        \centering
        \includegraphics[width=\linewidth]{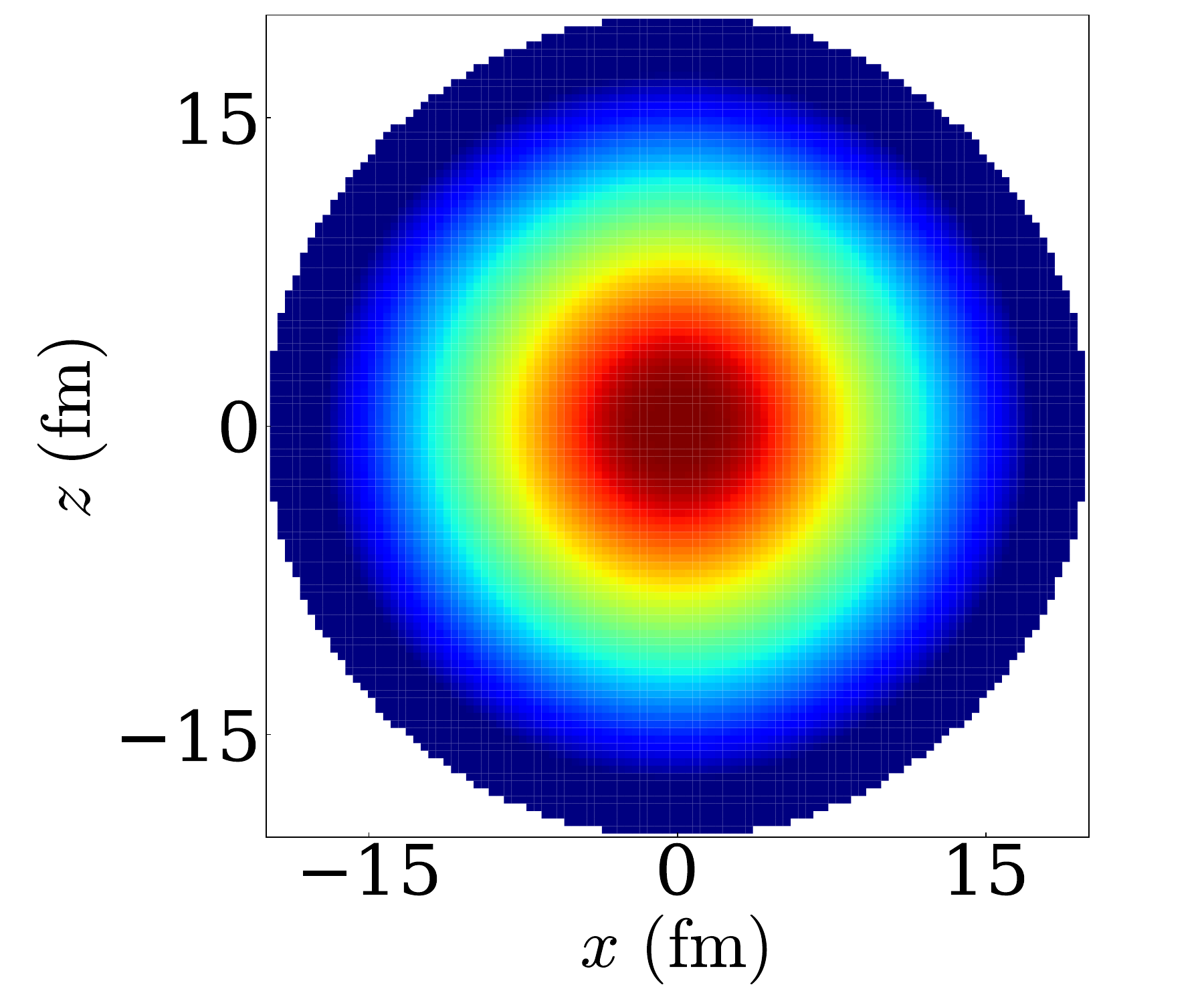}
        \caption{$^{28}$Ne}
    \end{subfigure}
    \hfill
    \begin{subfigure}[b]{0.17\textwidth}
        \centering
        \includegraphics[width=\linewidth]{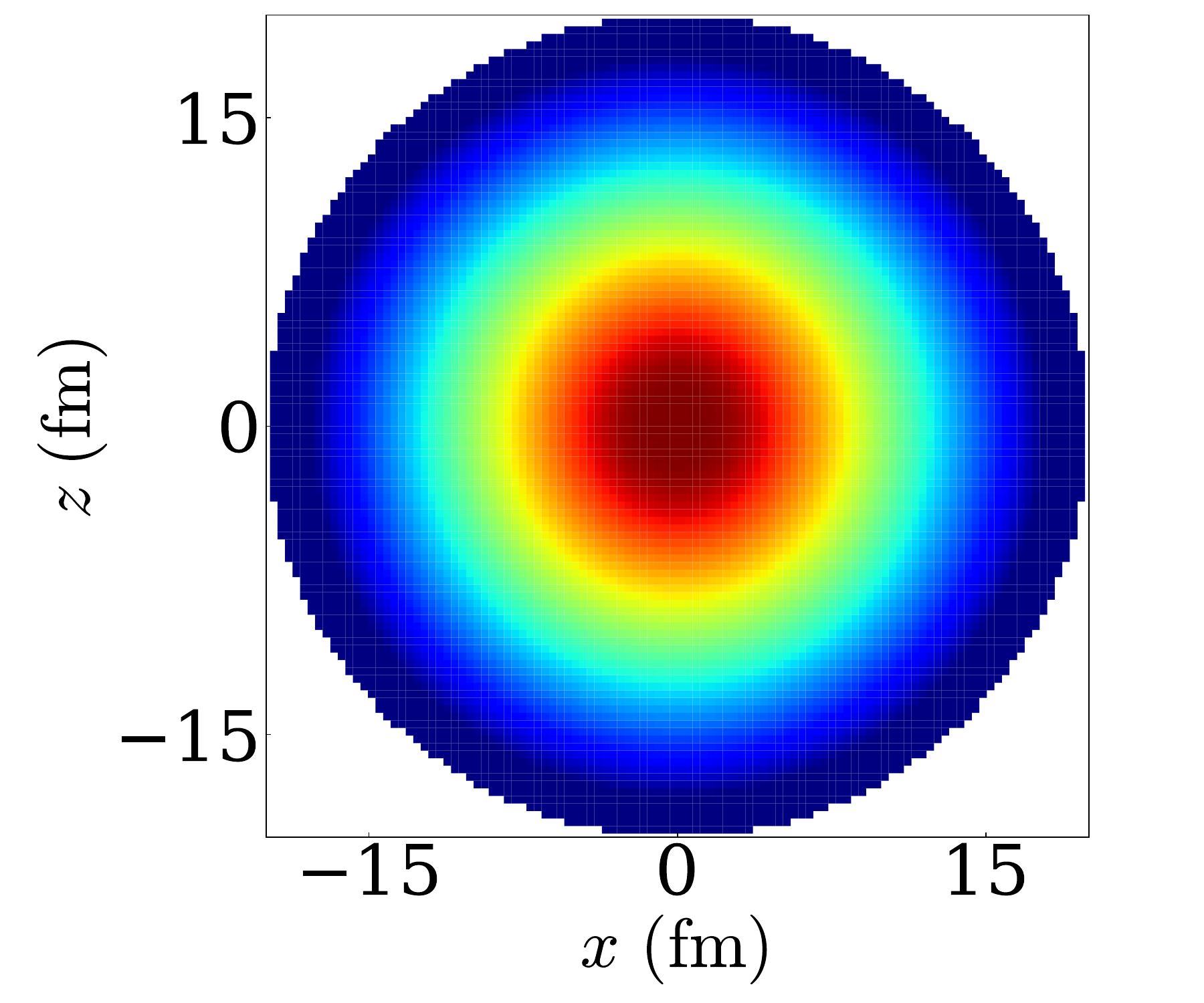}
        \caption{$^{29}$Ne}
    \end{subfigure}
    \hfill
    \begin{subfigure}[b]{0.17\textwidth}
        \centering
        \includegraphics[width=\linewidth]{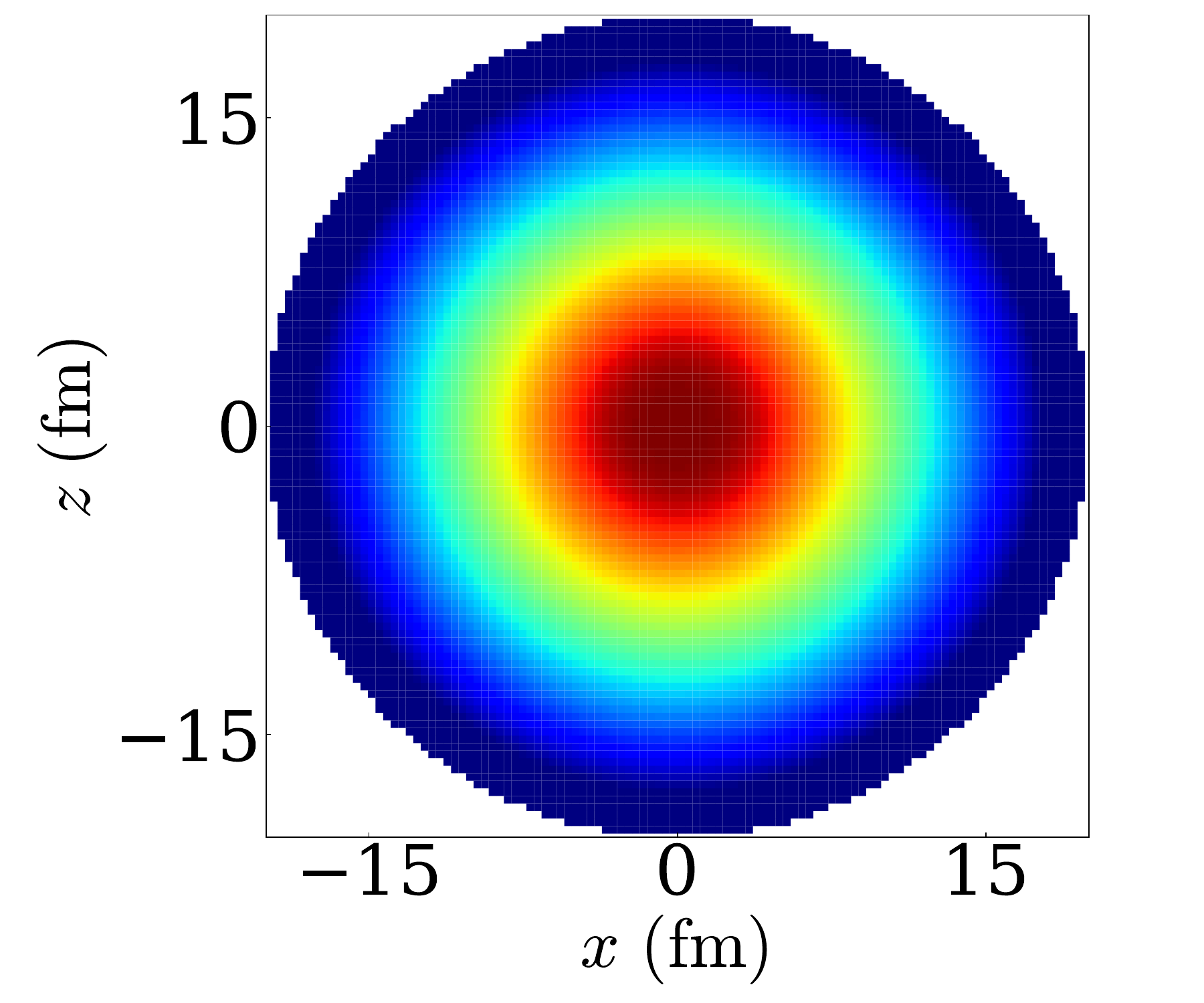}
        \caption{$^{30}$Ne}
    \end{subfigure}
    \hfill
    \begin{subfigure}[b]{0.17\textwidth}
        \centering
        \includegraphics[width=\linewidth]{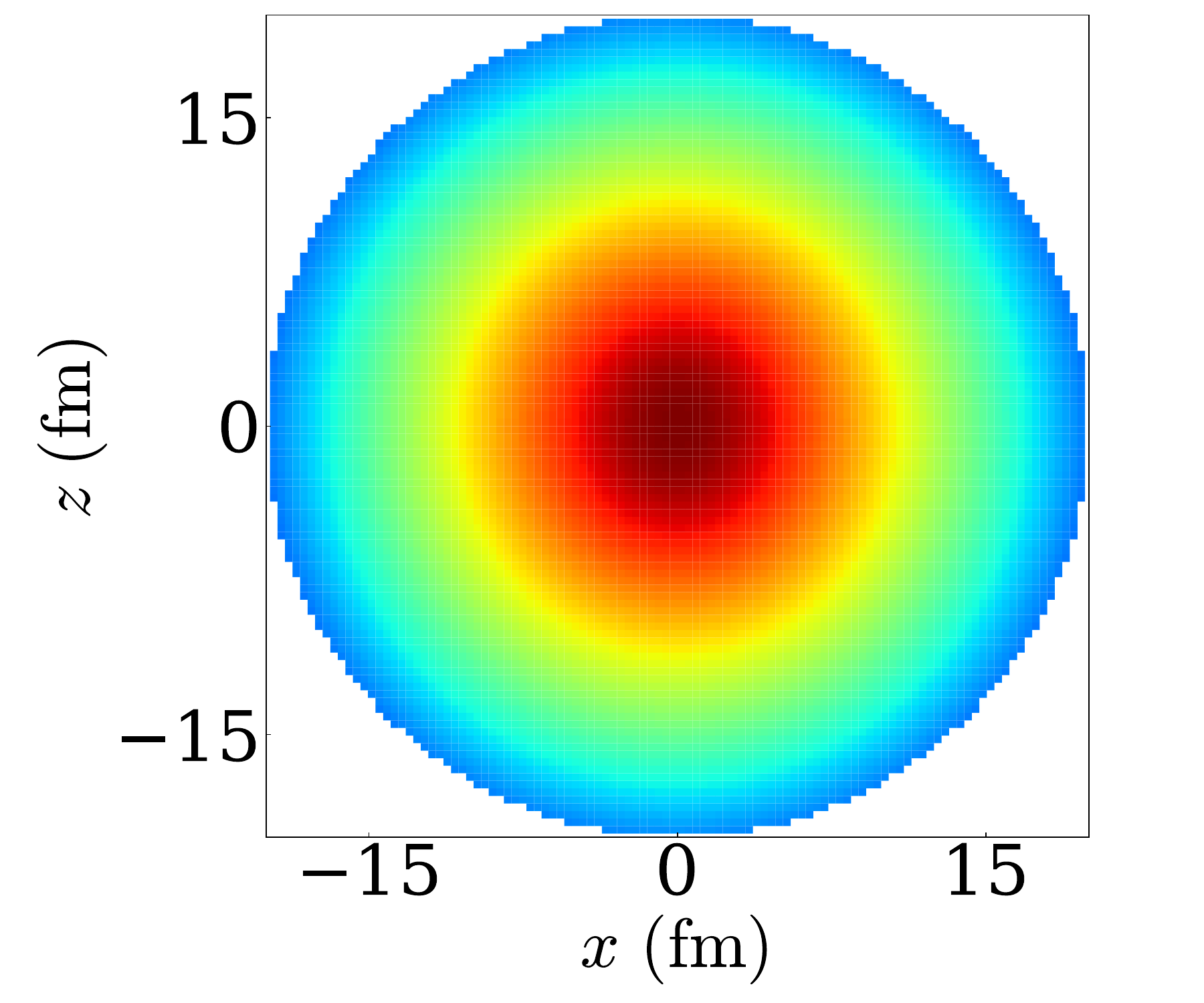}
        \caption{$^{31}$Ne}
    \end{subfigure}
    \hfill
    \begin{subfigure}[b]{0.17\textwidth}
        \centering
        \includegraphics[width=\linewidth]{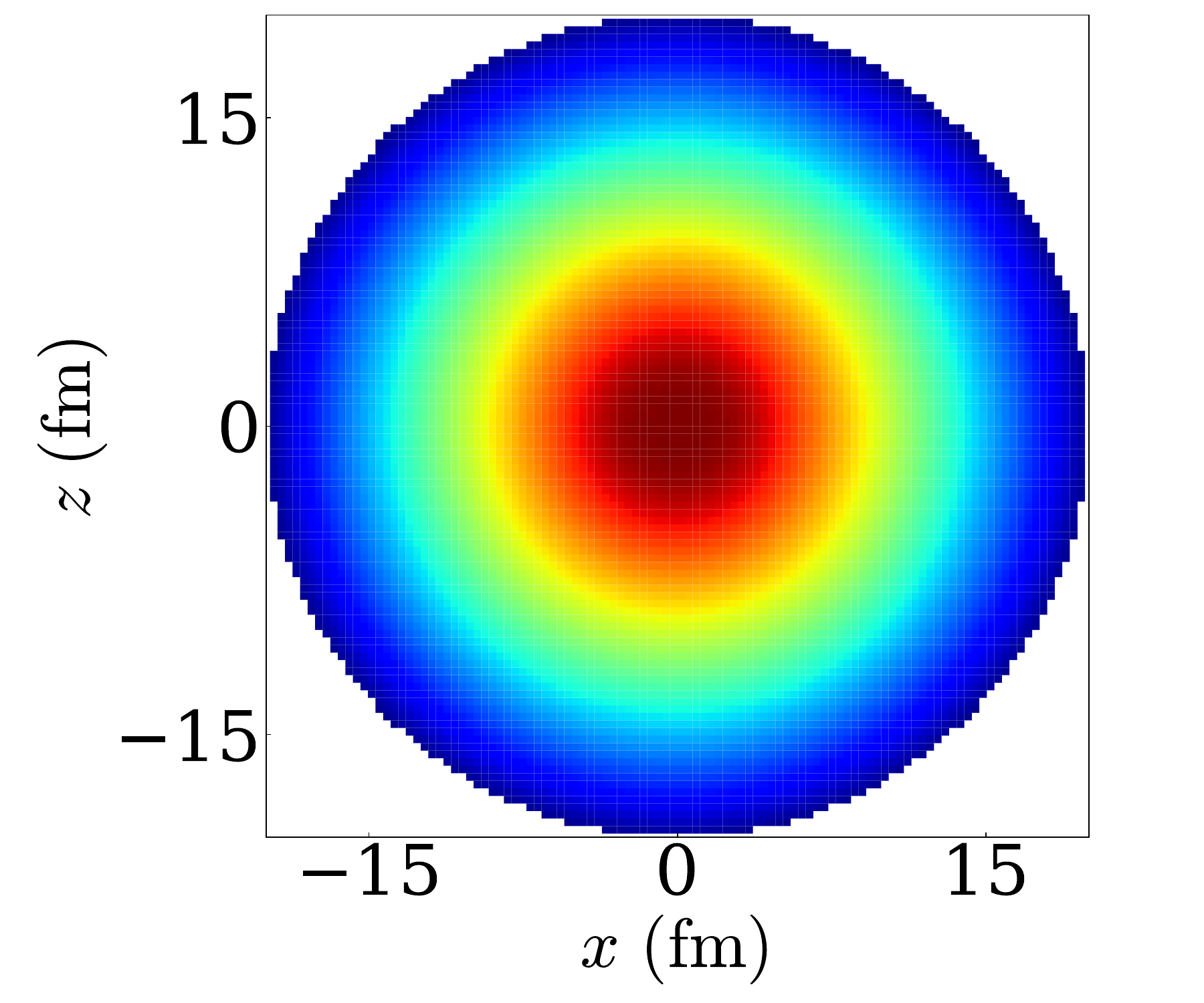}
        \caption{$^{32}$Ne}
    \end{subfigure}
    \hfill
    \begin{subfigure}[b]{0.05\textwidth}
        \centering
        \raisebox{0.8cm}{\includegraphics[height=0.10\textheight]{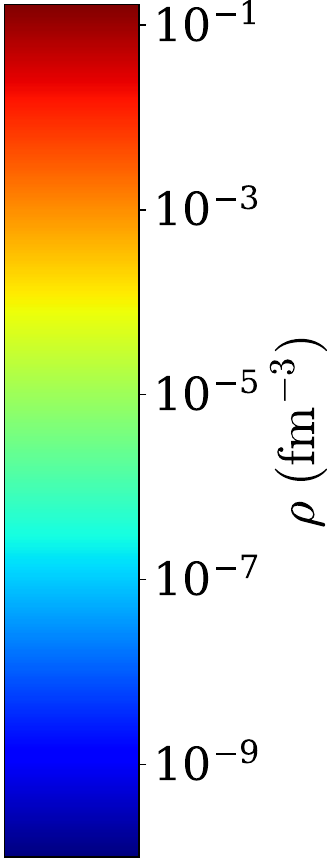}}
    \end{subfigure}
    
     \begin{subfigure}[b]{0.17\textwidth}
        \centering
        \includegraphics[width=\linewidth]{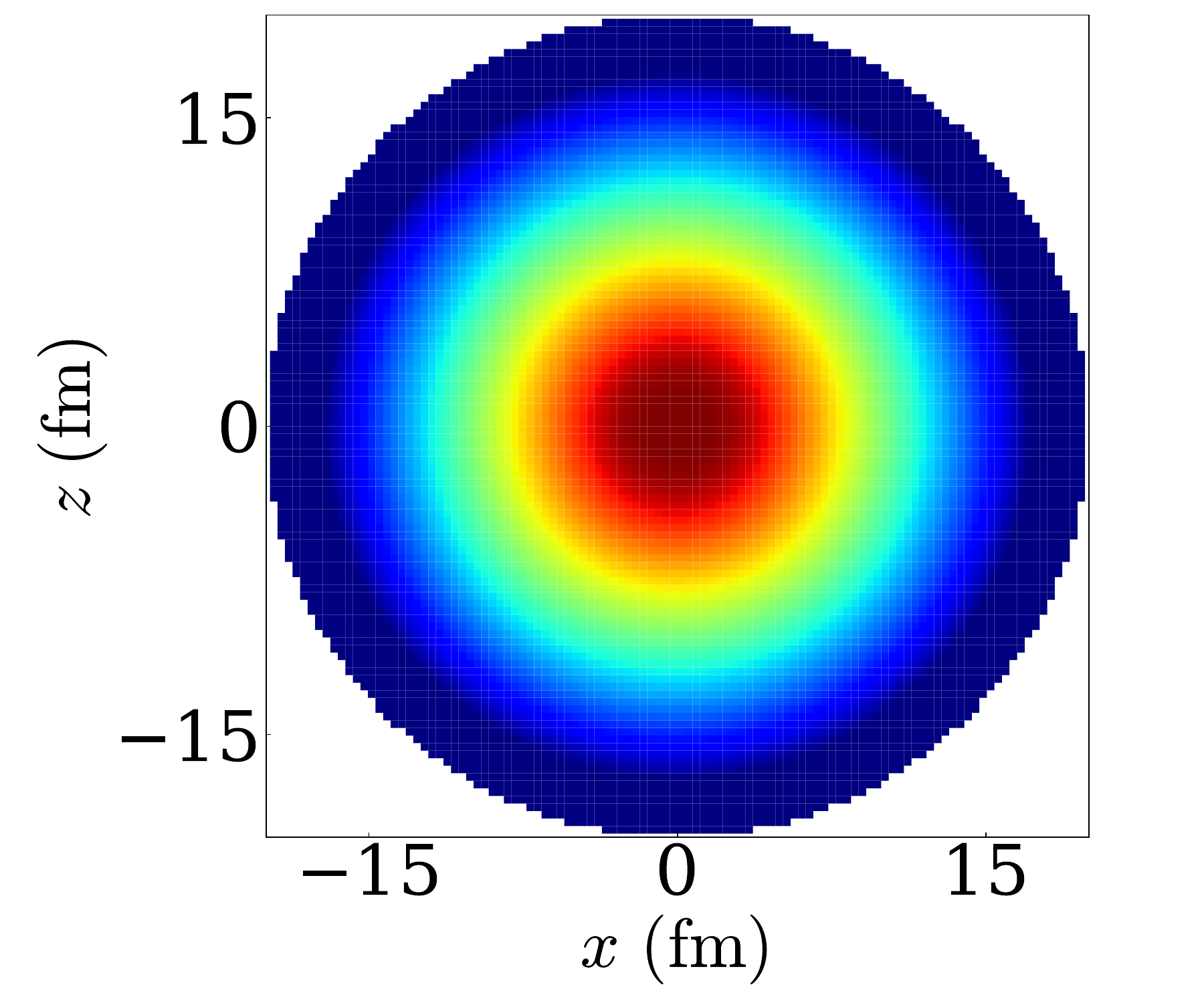}
        \caption{$^{28}$Ne}
    \end{subfigure}
    \hfill
    \begin{subfigure}[b]{0.17\textwidth}
        \centering
        \includegraphics[width=\linewidth]{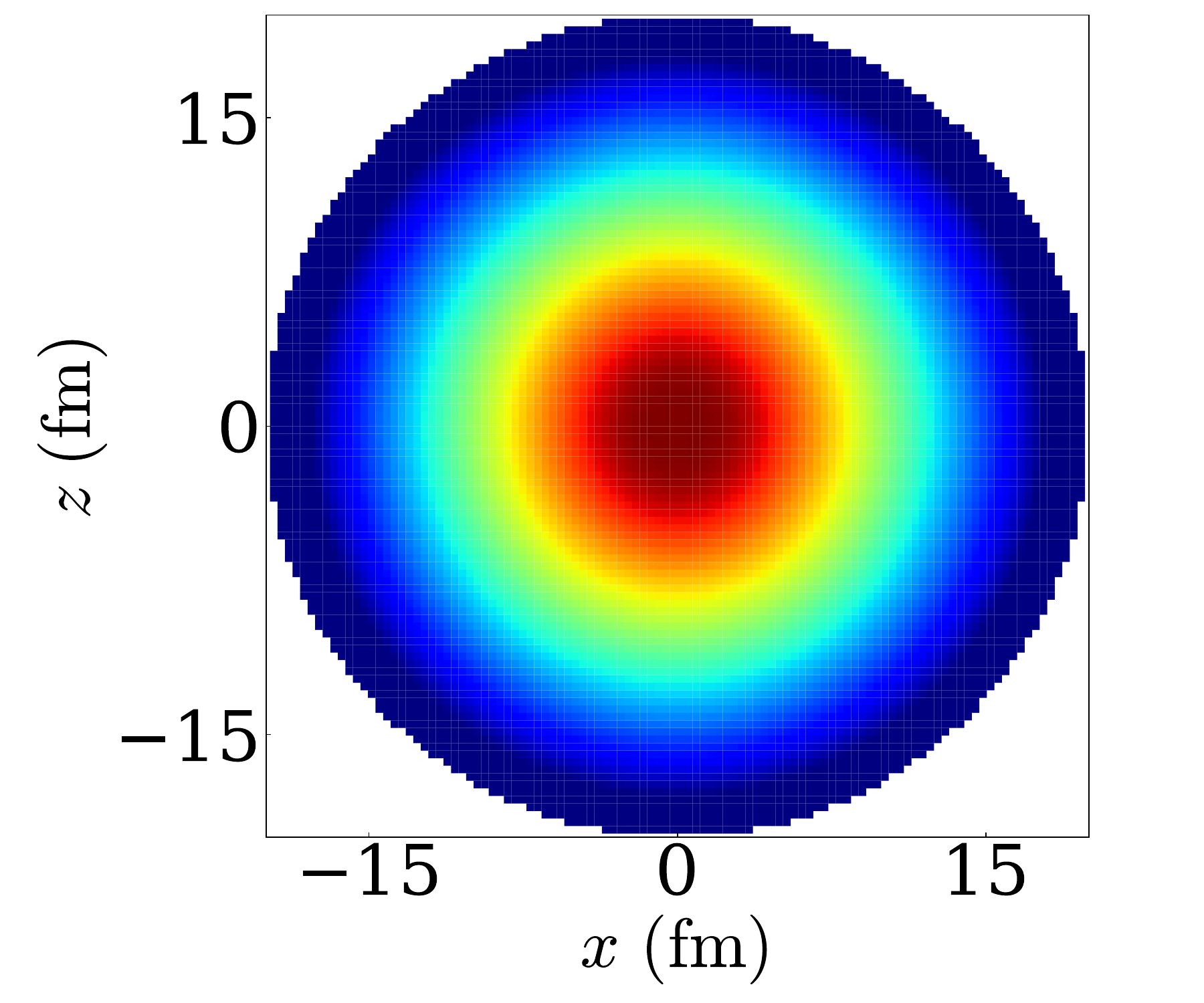}
        \caption{$^{29}$Ne}
    \end{subfigure}
    \hfill
    \begin{subfigure}[b]{0.17\textwidth}
        \centering
        \includegraphics[width=\linewidth]{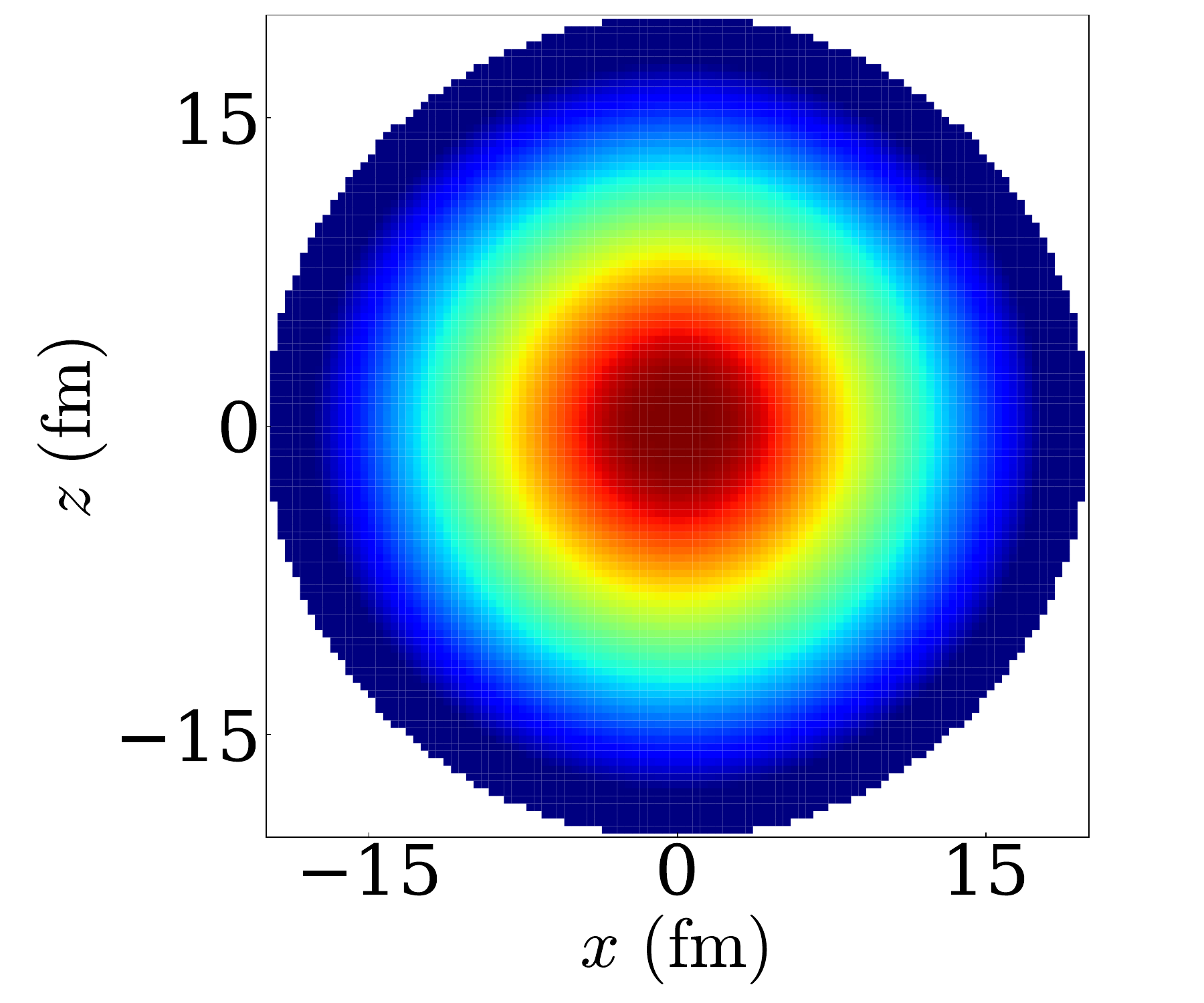}
        \caption{$^{30}$Ne}
    \end{subfigure}
    \hfill
    \begin{subfigure}[b]{0.17\textwidth}
        \centering
        \includegraphics[width=\linewidth]{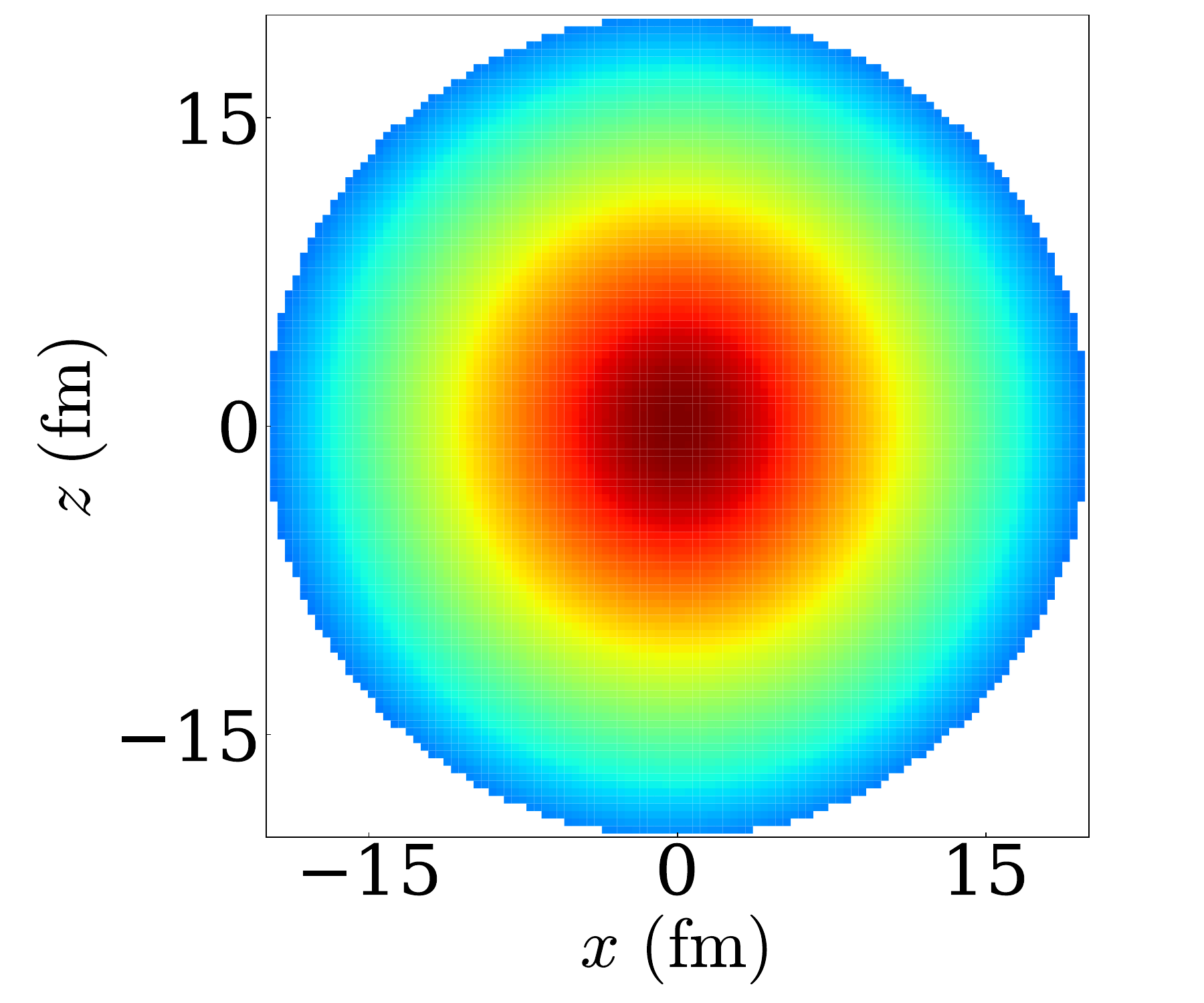}
        \caption{$^{31}$Ne}
    \end{subfigure}
    \hfill
    \begin{subfigure}[b]{0.17\textwidth}
        \centering
        \includegraphics[width=\linewidth]{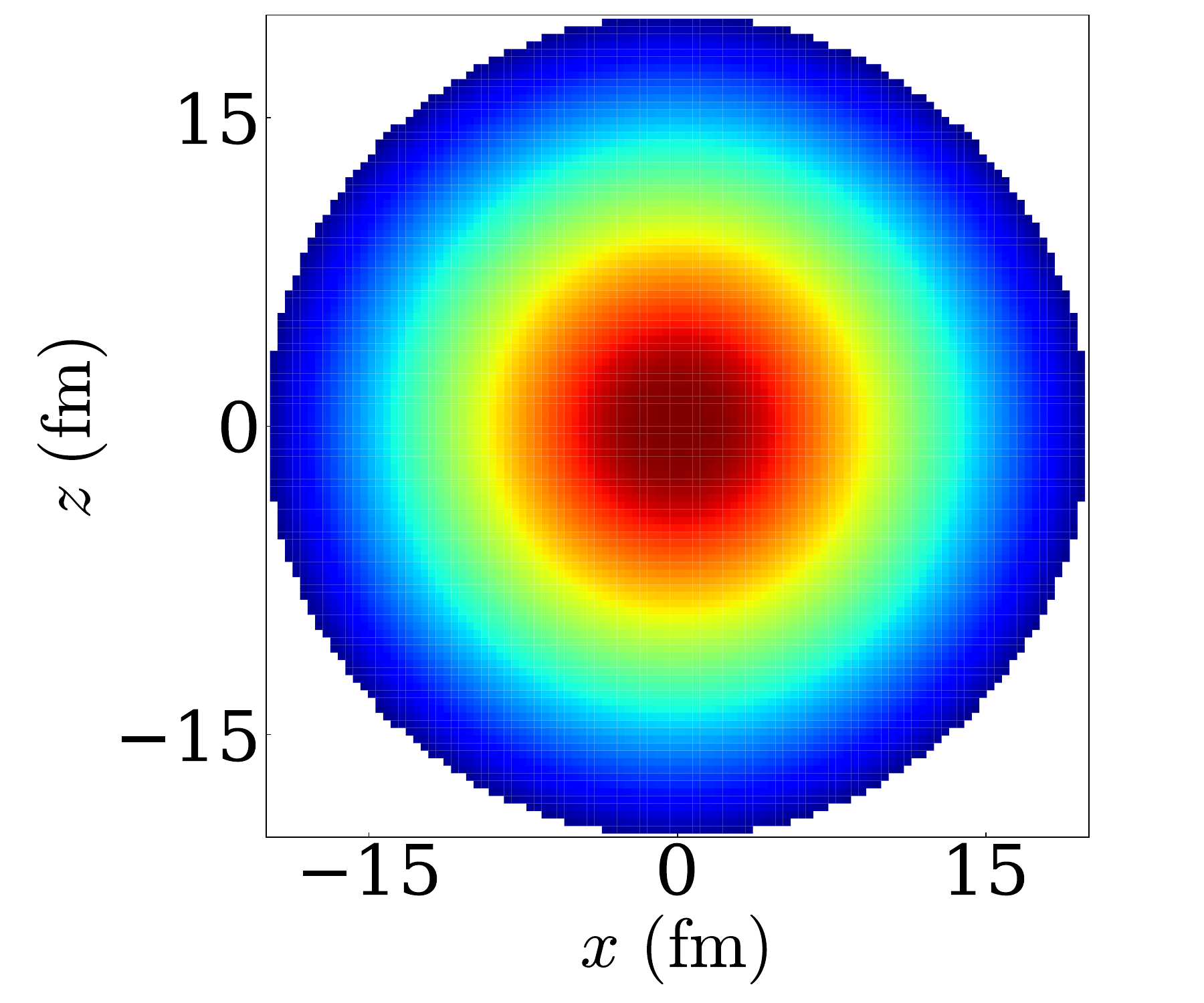}
        \caption{$^{32}$Ne}
    \end{subfigure}
    \hfill
    \begin{subfigure}[b]{0.05\textwidth}
        \centering
        \raisebox{0.8cm}{\includegraphics[height=0.10\textheight]{colorbar_WS.pdf}}
    \end{subfigure}

    \caption{(Color online) Matter density distributions of $^{28-32}\mathrm{Ne}$ in the $x$-$z$ plane obtained from the WS calculation when $\rho_0$ is taken as the density, respectively, \blu{at $r=0$ (upper) and $\rho_0$ at the maximum density (lower).} The corresponding values of $\rho_0$, $a$, $R_0$, $A_{WS}$, and the ratios $I_2/(I_1+I_2)$ are summarized in Table~\ref{tab:ws_fit_r0}.}
    \label{fig:density_xz_WS_at0}
\end{figure*}

\red{The fitted radius parameter $R_0$ shows a reduction for $^{31}$Ne under the same logarithmic tail fit and normalization constraints. In the present manuscript we interpret this decrease as a supporting consequence of the anomalously large diffuseness, not as an independent primary criterion, because $R_0$ is more sensitive to the fitting scheme and normalization choice.} The isotopic evolution of the diffuseness therefore reveals a clear and localized anomaly at $^{31}$Ne, consistent with the microscopic density analysis presented in Sec.~II. \blu{Figure~\ref{fig:density_xz_WS_at0} displays} the corresponding WS matter densities in the intrinsic frame.

\begin{table}[!t] 
\caption{Deformed WS fitting results for $^{28\text{-}32}$Ne, where the normalization parameter $\rho_0$ is taken as the central density $\rho(r=0)$ \blu{(upper) and  the maximum density $\rho_{\max}$ (lower)} of the DRHBc distribution. The extracted diffuseness parameter $a$ and the tail fraction $I_2/(I_1+I_2)$ are listed.}
\label{tab:ws_fit_r0}
\centering
\begin{tabular}{cccccc}
\hline\hline
Nucleus
& $^{28}$Ne & $^{29}$Ne & $^{30}$Ne & $^{31}$Ne & $^{32}$Ne \\
\hline
$\rho_0$ (fm$^{-3}$)
& 0.165 & 0.160 & 0.153 & 0.155 & 0.151 \\
$a$ (fm)
& 0.67 & 0.70 & 0.69 & 1.09 & 0.81 \\
$R_0$ (fm)
& 3.00 & 3.05 & 3.17 & 2.56 & 3.11 \\
$I_2/(I_1+I_2)$
& 0.51 & 0.53 & 0.50 & 0.78 & 0.58 \\
$A_{WS}$ & 27.86 & 28.92 & 29.97 & 30.86 & 31.81 \\
\hline\hline

\hline\hline
Nucleus
& $^{28}$Ne & $^{29}$Ne & $^{30}$Ne & $^{31}$Ne & $^{32}$Ne \\
\hline
$\rho_0$ (fm$^{-3}$)
& 0.169 & 0.164 & 0.158 & 0.161 & 0.156 \\
$a$ (fm)
& 0.67 & 0.70 & 0.69 & 1.09 & 0.81 \\
$R_0$ (fm)
& 2.97 & 3.02 & 3.12 & 2.50 & 3.07 \\
$I_2/(I_1+I_2)$
& 0.52 & 0.53 & 0.51 & 0.78 & 0.59 \\
$A_{WS}$ & 27.88 & 28.97 & 29.82 & 30.83 & 31.94 \\
\hline\hline

\end{tabular}
\end{table}

\FloatBarrier

\subsection{Formulation of a phenomenological halo criterion}

Based on the above observations, we propose the following phenomenological criterion for identifying halo nuclei in medium--mass systems:
(i) \red{The primary signature is an anomalously large surface diffuseness parameter $a$ extracted from the deformed WS fit relative to neighboring isotopes.} (ii) \red{A reduced fitted radius parameter $R_0$ may accompany the large diffuseness under the chosen normalization and logarithmic tail-fitting constraint, but it is treated here only as a supporting consequence rather than as an independent criterion.}

These conditions are satisfied unambiguously for $^{31}$Ne, whereas $^{29}$Ne does not fulfill either of them. For $^{32}$Ne, the results indicate an intermediate situation, suggesting that its extended structure may originate from a thick neutron skin rather than a genuine halo. The diffuseness--based criterion developed in this section therefore provides a locally defined and quantitative measure of halo formation in coordinate space. In the next section, we examine how the same extended density tails manifest themselves in momentum space through nuclear form factors within the Helm model framework.

\section{Helm-model analysis and comparison with the Woods--Saxon description}

While the WS fitting in Sec.~III provides
a local and coordinate-space measure of the surface diffuseness,
it remains instructive to examine how the same density distributions
manifest themselves in momentum space.
For this purpose, we employ the Helm model, which offers
a compact and geometrically transparent representation of
nuclear densities in terms of form factors. 
Unlike the WS parametrization, which directly probes the radial
falloff of the density profile,
the Helm model characterizes the density through
its Fourier transform and thus provides complementary information
on spatial extension. The nuclear form factor is defined as the Fourier transform
of the density distribution,
\begin{equation}
F(q) = \int e^{i\mathbf{q}\cdot\mathbf{r}} \rho(\mathbf{r}) d^3r.
\end{equation}

For spherical symmetry, this reduces to
\begin{equation}
F(q) = 4\pi \int_0^\infty j_0(qr)\,\rho(r)\, r^2 dr ,
\end{equation}
where $j_0$ is the spherical Bessel function. In the Helm model, the density is expressed as the convolution
of a hard sphere of diffraction radius $R_0$ with a Gaussian
of width $\sigma$ (surface thickness parameter).
The corresponding form factor reads
\begin{equation}
F(q) =
\frac{3j_1(qR_0)}{qR_0}
\exp\!\left(-\frac{1}{2} q^2 \sigma^2\right).
\label{eq:Helm_form}
\end{equation}

\red{The diffraction radius $R_0$ is extracted from the first zero $q_1$ of the DRHBc form factor,}
\begin{equation}
R_0 = \frac{4.49341}{q_1},
\end{equation}
while the surface thickness parameter $\sigma$ is determined
from the behavior near the first maximum of the form factor as
\begin{equation}
    \sigma^2 = \frac{2}{q_m^2} \ln \left( \frac{3 R_0^2 j_1(q_m R_0)}{R_0 q_m F(q_m)} \right)
\end{equation}
where $q_m$ is the first maximum of the form factor obtained using the DRHBc approach. The Helm-model rms radius is given by
\begin{equation}
R_{\mathrm{rms}}^{(H)} =
\sqrt{\frac{3}{5}(R_0^2 + 5\sigma^2)}.
\end{equation}
\red{The Helm parameters quoted below are obtained directly from the DRHBc form factor; the WS parametrization is used only as a qualitative cross-check.} A useful diagnostic quantity is the difference
\begin{equation}
\Delta R_{\mathrm{rms}} =
R_{\mathrm{rms}}^{\mathrm{DRHBc}}
-
R_{\mathrm{rms}}^{(H)} ,
\end{equation}
which quantifies how much of the microscopic radius
cannot be reproduced by a simple Helm parametrization.

\subsection{Comparison between Helm and Woods--Saxon descriptions}

Although both the WS and Helm models provide effective
parametrizations of nuclear densities,
they differ fundamentally in their sensitivity
to halo-like structures, as shown in Fig.~\ref{fig:ws_vs_helm}. The WS diffuseness parameter $a$ directly controls
the exponential falloff of the density in coordinate space.
Therefore, it is highly sensitive to weakly bound
valence nucleons that generate extended tails.
In contrast, the Helm model describes the density
through its form factor and effectively smooths
the radial distribution by Gaussian folding. \red{The isotopic variation of $\Delta R_{\mathrm{rms}}$ is significantly smaller than that of the diffuseness parameter $a$, showing that the Helm representation is less sensitive to the extremely dilute asymptotic tail.}
As a result, very dilute long-range tails may have
only a moderate impact on the extracted Helm parameters.

This difference is clearly reflected in the Ne isotopes.
For $^{31}$Ne, the WS analysis yields a pronounced enhancement
of the diffuseness parameter $a \simeq 1.09$~fm,
indicating a slow radial decay.
However, the Helm-model parameters $R_0$ and $\sigma$
do not exhibit an equally dramatic anomaly.
Instead, the primary signal appears in the difference
$\Delta R_{\mathrm{rms}}$,
which becomes largest for $^{31}$Ne when spherical symmetry
is assumed.

In other words, while the WS parametrization detects
halo formation through an anomalous surface thickness,
the Helm model reveals it indirectly through
the inability of a simple folded geometry
to reproduce the full rms radius.

\subsection{Matter versus charge form factors}

An important consistency check is obtained by comparing
matter and charge form factors.
Since halo formation in neutron-rich nuclei is driven
primarily by weakly bound neutrons,
the charge distribution is expected to remain largely unchanged. Indeed, the Helm parameters extracted from the proton
density distributions show only minor isotopic variation.
The difference
\begin{equation}
\Delta R^{(H)} =
R_{\mathrm{rms,matter}}^{(H)}
-
R_{\mathrm{rms,charge}}^{(H)}
\end{equation}
does not display a sharp enhancement
except for the neutron-rich edge of the chain as shown in Table~\ref{tab:helm_comparison}.
This confirms that the extended structure in $^{31}$Ne
originates predominantly from the neutron density.

\begin{figure}[t] 
  \centering
  \includegraphics[width=0.45\linewidth]{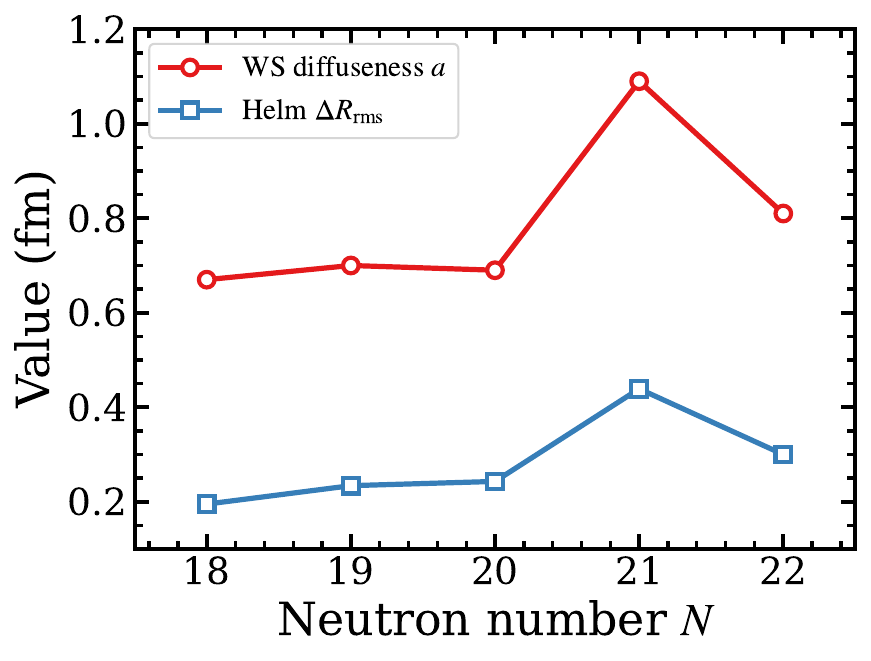}
  \caption{(Color online)
  Comparison between the Woods--Saxon diffuseness parameter $a$
  and the Helm-model indicator $\Delta R_{\mathrm{rms}}$
  for $^{28\text{--}32}$Ne.
  The WS diffuseness exhibits a pronounced enhancement
  at $N=21$ ($^{31}$Ne),
  while the Helm parameter reflects the anomaly
  indirectly through the increased mismatch
  between the microscopic and folded radii.
  This demonstrates the higher sensitivity of the WS
  parametrization to weakly bound halo structures.}
  \label{fig:ws_vs_helm}
\end{figure}

\begin{table*}[t] 
\caption{Comparison of Helm-model results extracted from
matter and charge form factors for $^{28\text{--}32}$Ne.
The Helm rms radii $R_{\mathrm{rms}}^{(H)}$ and the differences
$\Delta R_{\mathrm{rms}}
= R_{\mathrm{rms}}^{(\mathrm{DRHBc})}
- R_{\mathrm{rms}}^{(H)}$
are listed separately for matter and charge distributions.}
\label{tab:helm_comparison}
\centering
\begin{tabular}{ccccccc}
\hline\hline
 & \multicolumn{3}{c}{Matter} 
 & \multicolumn{3}{c}{Charge} \\
\cline{2-4} \cline{5-7}
Nucleus 
& $R_{\mathrm{rms}}^{(H)}$ 
& $R_{\mathrm{rms}}^{(\mathrm{DRHBc})}$ 
& $\Delta R_{\mathrm{rms}}$
& $R_{\mathrm{rms}}^{(H)}$
& $R_{\mathrm{rms}}^{(\mathrm{DRHBc})}$
& $\Delta R_{\mathrm{rms}}$ \\
\hline
$^{28}$Ne 
& 2.972 & 3.167 & 0.195 
& 2.970 & 2.859 & -0.111 \\
$^{29}$Ne 
& 3.000 & 3.234 & 0.234 
& 2.989 & 2.891 & -0.098 \\
$^{30}$Ne 
& 3.045 & 3.288 & 0.243 
& 3.015 & 2.921 & -0.094 \\
$^{31}$Ne 
& 3.060 & 3.499 & 0.439 
& 3.022 & 2.927 & -0.095 \\
$^{32}$Ne 
& 3.130 & 3.430 & 0.300 
& 3.053 & 2.947 & -0.106 \\
\hline\hline
\end{tabular}
\end{table*}

Table~\ref{tab:helm_comparison} compares the Helm-model
results obtained from matter and charge form factors.
A pronounced enhancement of
$\Delta R_{\mathrm{rms}}$ is observed for $^{31}$Ne
in the matter channel,
whereas the corresponding charge values remain nearly constant
along the isotopic chain.
This clearly demonstrates that the extended structure in
$^{31}$Ne originates predominantly from the neutron density.
The absence of a comparable anomaly in the charge channel
confirms that the halo phenomenon is neutron-driven.

A direct comparison between the WS diffuseness parameter $a$
and the Helm-model indicator $\Delta R_{\mathrm{rms}}$
is shown in Fig.~\ref{fig:ws_vs_helm}.
While both quantities signal an anomaly at $^{31}$Ne,
their physical origins are different.

The WS diffuseness probes the local radial slope of the
density distribution and therefore responds directly
to the slow exponential decay associated with weak binding.
In contrast, the Helm model characterizes the density
through a folded geometric representation in momentum space.
As a consequence, the extended dilute tail of a halo nucleus
does not strongly modify the Helm surface parameter $\sigma$,
but instead appears as an increased discrepancy
$\Delta R_{\mathrm{rms}}$. This comparison clearly shows that the WS parametrization
is more sensitive to halo formation in coordinate space,
whereas the Helm analysis provides a complementary,
momentum-space perspective.

\subsection{Helm-model formalism with deformation}

For axially deformed nuclei, the density distribution
can be expanded in multipoles as
\begin{equation}
\rho(r,\theta)
=
\sum_{\lambda=0,2,4,\dots}
\rho_{\lambda}(r)\,
P_{\lambda}(\cos\theta),
\label{eq:density_multipole}
\end{equation}
where $P_{\lambda}$ are Legendre polynomials. The form factor becomes orientation dependent,
\begin{equation}
F(q,\theta')
=
\sum_{\ell=0}^{\infty}
F_{\ell}(q)\,
P_{\ell}(\cos\theta'),
\label{eq:F_multipole}
\end{equation}
with
\begin{equation}
F_{\ell}(q)
=
4\pi i^{\ell}
\int_0^\infty
j_{\ell}(qr)\,
\rho_{\ell}(r)\,
r^2 dr .
\label{eq:F_l}
\end{equation}

Since experimental observables are insensitive to the
intrinsic orientation of the nucleus,
an orientation average of the squared form factor
is performed,
\begin{equation}
\left\langle |F(q)|^2 \right\rangle
=
\frac{1}{2}
\int_{-1}^{1}
|F(q,\theta')|^2 \,
d(\cos\theta')
=
\sum_{\ell=0}^{\infty}
(2\ell+1)
|F_{\ell}(q)|^2 .
\label{eq:orientation_avg}
\end{equation}

The effective Helm parameters $R_0$ and $\sigma$
are then extracted by fitting
$\sqrt{\langle |F(q)|^2 \rangle}$
with Eq.~(\ref{eq:Helm_form}).

In brief, the WS parametrization probes the local radial slope
of the density in coordinate space,
whereas the Helm model characterizes the density
through its Fourier-transformed geometry.
Deformation primarily modifies the effective surface thickness
$\sigma$ via orientation averaging,
while weak binding and continuum coupling
manifest themselves more directly
in the asymptotic behavior of the density,
reflected in the diffuseness parameter $a$.

When deformation effects are included,
the Helm surface parameter $\sigma$
increases moderately for the deformed isotopes.
However, even after incorporating deformation,
the WS diffuseness remains significantly larger
for $^{31}$Ne than for its neighbors. \red{This indicates that quadrupole deformation alone cannot generate the extended density tail; the halo signature instead points to weak binding and continuum coupling effects \cite{Mizutori2000}.}

\begin{table}[t]
\caption{Helm-model parameters extracted from the matter form factors
of $^{28\text{--}32}$Ne including deformation effects via orientation
averaging.
The diffraction radius $R_0$, surface thickness $\sigma$,
Helm rms radius $R_{\mathrm{rms}}^{(H)}$,
and the microscopic DRHBc rms radius
$R_{\mathrm{rms}}^{(\mathrm{DRHBc})}$ are listed.}
\label{tab:helm_deformed}
\centering
\begin{tabular}{ccccc}
\hline\hline
Nucleus & $R_0$ (fm) & $\sigma$ (fm) 
& $R_{\mathrm{rms}}^{(H)}$ (fm) 
& $R_{\mathrm{rms}}^{(\mathrm{DRHBc})}$ (fm) \\
\hline
$^{28}$Ne & 3.30 & 0.87 & 2.972 & 3.167 \\
$^{29}$Ne & 3.38 & 0.98 & 3.122 & 3.234 \\
$^{30}$Ne & 3.46 & 0.83 & 3.045 & 3.288 \\
$^{31}$Ne & 3.45 & 1.01 & 3.193 & 3.499 \\
$^{32}$Ne & 3.52 & 0.89 & 3.130 & 3.430 \\
\hline\hline
\end{tabular}
\end{table}

Table~\ref{tab:helm_deformed} summarizes the Helm parameters
extracted from the matter form factors when deformation effects
are included through orientation averaging.
Compared with the spherical approximation,
the surface thickness parameter $\sigma$
increases noticeably for the deformed isotopes
$^{29}$Ne and $^{31}$Ne,
reflecting the geometric smearing induced by quadrupole deformation. For $^{31}$Ne, the deformation-inclusive Helm analysis yields
$\sigma \simeq 1.01$~fm and
$R_{\mathrm{rms}}^{(H)} = 3.193$~fm,
which is significantly closer to the microscopic
DRHBc value $R_{\mathrm{rms}}^{(\mathrm{DRHBc})}=3.499$~fm
than in the spherical treatment.
Nevertheless, a substantial difference remains,
indicating that deformation alone cannot account for
the full spatial extension.
This residual mismatch originates from the weak binding
and the extended neutron density tail,
which is more directly captured by the Woods--Saxon
diffuseness parameter $a$ discussed in Sec.~III.

\section{Interaction cross sections and halo robustness}

\subsection{Glauber MOL framework}

The interaction cross sections $\sigma_I$ are evaluated using the
Glauber model within the modified optical limit (MOL)
prescription~\cite{AbuIbrahim2000}, including a finite-range \blu{nucleon-nucleon (NN)} profile
and the Fermi-motion averaged effective NN total cross section
following Ref.~\cite{Takechi2009}.
In the present energy region ($E\simeq 240$ MeV/nucleon),
the contribution of inelastic excitation to the reaction cross section is small,
so that the experimentally reported interaction cross section can be identified with
the reaction cross section to good accuracy~\cite{Takechi2012}.
Accordingly, throughout this section we use the notation
\begin{equation}
\sigma_I \simeq \sigma_R,
\end{equation}
and treat the Glauber-calculated reaction observable as $\sigma_I$. \red{This structure-to-reaction perspective is consistent with earlier reaction analyses of $^{31}$Ne and with recent unified DRHBc+Glauber applications to halo nuclei in the C and Mg chains \cite{Urata2012,Wang2024EPJA,An2024PLB}.} Experimental interaction cross sections for $^{28\text{--}32}$Ne are taken from
Ref.~\cite{Takechi2012}.
The NN total-cross-section inputs are based on the parametrizations of
Norbury~\cite{Norbury2008} and Cugnon \textit{et al.}~\cite{Cugnon1996}.

\subsection{Glauber model formalism}

The interaction (reaction) observable is evaluated within the
Glauber multiple-scattering framework in the eikonal approximation~\cite{Glauber1959},
where the projectile--target scattering is described in impact-parameter space. The survival probability at a given impact parameter $b$ is written as
\begin{equation}
P_{\mathrm{surv}}(b)
=
|S(b)|^2
=
\exp\!\left[-2\,\mathrm{Im}\,\chi(b)\right],
\label{eq:survival}
\end{equation}
where $S(b)=\exp[i\chi(b)]$ is the eikonal $S$-matrix and $\chi(b)$ is the eikonal phase. The corresponding interaction probability is
\begin{equation}
P_{I}(b)
=
1 - P_{\mathrm{surv}}(b)
=
1 - \exp\!\left[-2\,\mathrm{Im}\,\chi(b)\right],
\label{eq:interaction_probability}
\end{equation}
and the total interaction cross section becomes~\cite{Glauber1959}
\begin{equation}
\sigma_I
=
2\pi
\int_0^{\infty}
b\, db \,
\left[
1 - \exp\!\left(-2\,\mathrm{Im}\,\chi(b)\right)
\right].
\label{eq:sigmaI_final}
\end{equation}

In the optical-limit approximation (OLA)~\cite{Glauber1959},
the eikonal phase is expressed in terms of
the projectile and target thickness functions,
\begin{equation}
\chi_{\mathrm{OLA}}(b)
=
i
\int d^2s
\int d^2t \,
\Gamma(b+s-t)\,
\rho_{P,z}(s)\,
\rho_{T,z}(t),
\label{eq:chi_ola}
\end{equation}
where
\begin{equation}
\rho_{z}(s)
=
\int_{-\infty}^{+\infty}
\rho(s,z)\, dz
\label{eq:thickness}
\end{equation}
is the thickness function, and $\Gamma(b)$ is the nucleon--nucleon (NN)
profile function.

\subsection{NN-input dependence and robustness of the $^{31}$Ne anomaly}
\label{subsec:nn_robust}

\begin{figure*}[t]
  \centering
  \begin{minipage}{0.48\textwidth}
    \centering
    \includegraphics[width=\linewidth]{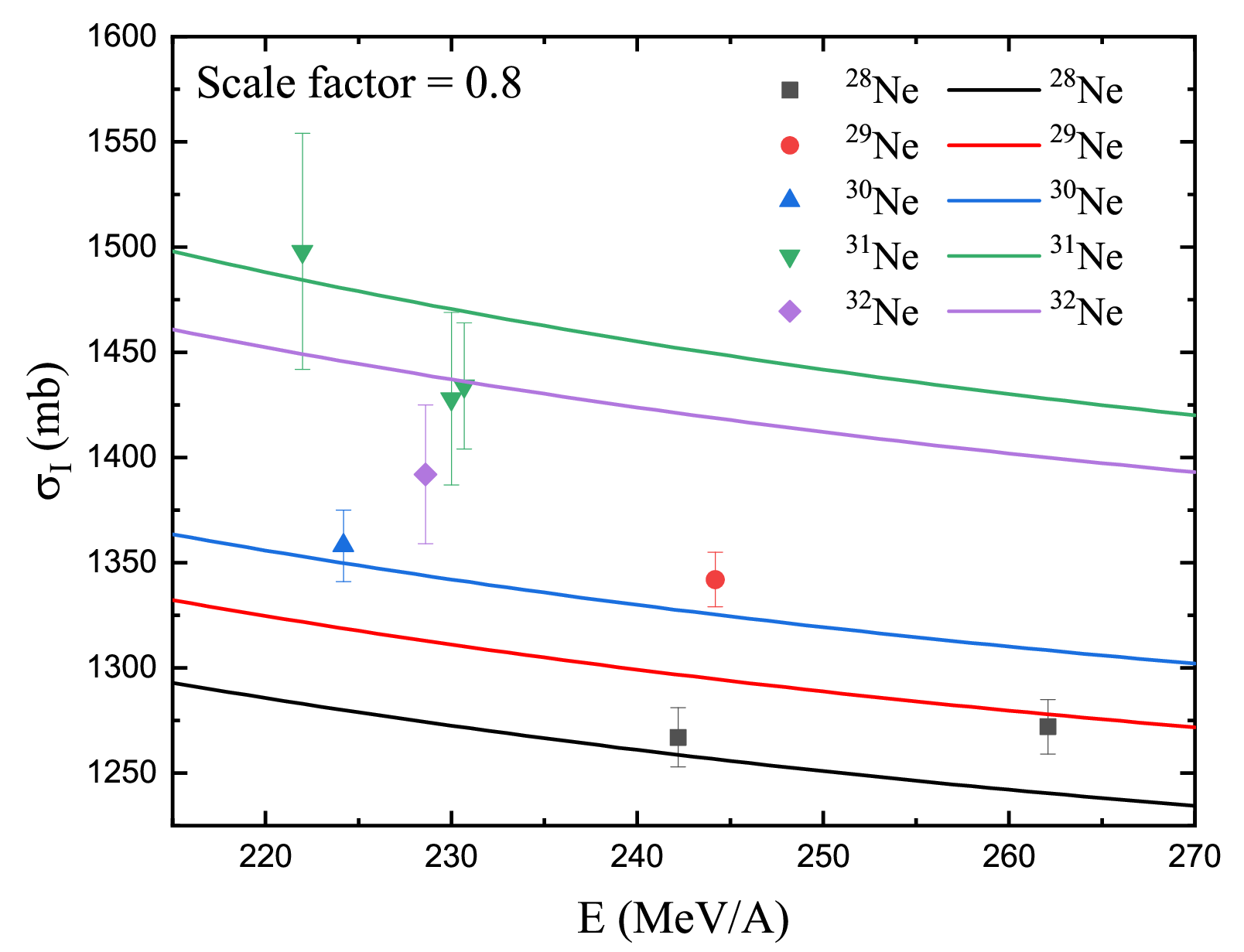}
    \vspace{-2mm}
    \textbf{(a)}
  \end{minipage}
  \hfill
  \begin{minipage}{0.48\textwidth}
    \centering
    \includegraphics[width=\linewidth]{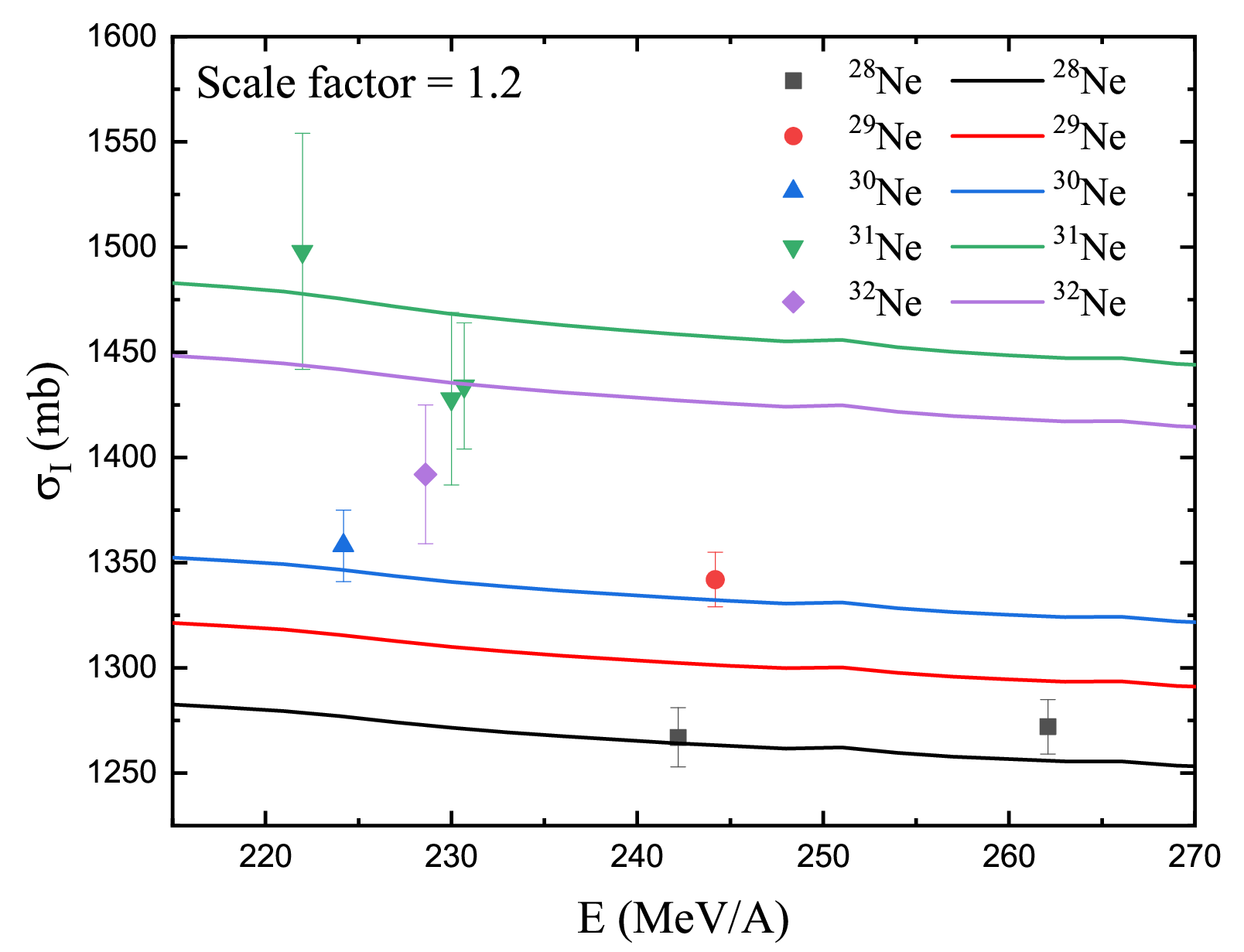}
    \vspace{-2mm}
    \textbf{(b)}
  \end{minipage}
  \caption{(Color online)
  Interaction cross sections $\sigma_I$ for
  $^{28\text{--}32}$Ne on a $^{12}$C target at
  $E\simeq 240$ MeV/nucleon calculated within the
  Glauber MOL[FM] framework~\cite{AbuIbrahim2000,Takechi2009}
  using two different NN total-cross-section parametrizations:
  (a) Norbury(2008)~\cite{Norbury2008} with the adopted global rescaling ($0.8\times$Norbury) and
  (b) Cugnon(1996)~\cite{Cugnon1996} with the adopted global rescaling ($1.2\times$Cugnon).
  Experimental points are taken from Ref.~\cite{Takechi2012}.
  \red{Although the absolute magnitude depends on the NN input, the relative enhancement at $^{31}$Ne is preserved, indicating that the isotopic anomaly reflects intrinsic nuclear structure rather than the particular NN prescription. The Coulomb correction is included in all calculations.}}
  \label{fig:sigma_combined}
\end{figure*}

Figure~\ref{fig:sigma_combined} displays the calculated interaction cross sections
$\sigma_I$ for $^{28\text{--}32}$Ne on a $^{12}$C target at energies close to
$E\simeq240$ MeV/nucleon.
Two different NN-input prescriptions are adopted:
the Norbury(2008) parametrization~\cite{Norbury2008} with a global factor $0.8$
and the Cugnon(1996) parametrization~\cite{Cugnon1996} with a global factor $1.2$.
Experimental interaction cross sections are taken from Ref.~\cite{Takechi2012}.
All calculations include the Coulomb correction.
For each NN prescription, a single global rescaling factor is chosen to bring the calculated absolute interaction-cross-section scale into reasonable overall agreement with the measured isotopic chain and is not readjusted isotope by isotope; the comparison therefore tests the robustness of the relative isotopic trend rather than a nucleus-by-nucleus renormalization.

\red{Although the absolute magnitude of $\sigma_I$ depends moderately on the adopted NN input and normalization, the relative isotopic trend remains stable. In particular, $^{31}$Ne consistently exhibits a clear enhancement compared with its neighboring isotopes under both NN prescriptions. This shows that the $^{31}$Ne anomaly is robust at the relative level across the NN prescriptions considered here.}

The NN profile function in \blu{Eq.~(\ref{eq:profile})} is taken in a Gaussian finite-range form~\cite{Takechi2009}
\begin{equation}
\Gamma(b)
=
\frac{1 - i\alpha}
{4\pi\beta}
\sigma_{NN}^{\mathrm{eff}}(E)
\exp\!\left(
-\frac{b^2}{2\beta}
\right),
\label{eq:profile}
\end{equation}
where $\sigma_{NN}^{\mathrm{eff}}$ is the effective NN total cross section,
$\alpha$ is the ratio of real to imaginary forward scattering amplitude,
and $\beta$ is the slope parameter related to the finite interaction range. To improve upon the linear OLA approximation, the modified optical limit (MOL) includes
higher-order multiple-scattering effects~\cite{AbuIbrahim2000}. The MOL phase can be written as
\begin{equation}
\chi_{\mathrm{MOL}}(b)
=
\frac{i}{2}
\int d^2s\, \rho_{P,z}(s)
\left[
1 - \exp
\left(
-\int d^2t\,
\rho_{T,z}(t)
\Gamma(b+s-t)
\right)
\right]
+ (P \leftrightarrow T),
\label{eq:chi_mol}
\end{equation}
with the thickness function $\rho_{z}(s)=\int \rho(s,z)\,dz$.

At intermediate energies, the NN total cross section depends strongly on energy.
To account for the Fermi motion of nucleons inside nuclei, an effective NN cross section
is introduced following Ref.~\cite{Takechi2009},
\begin{equation}
\sigma_{NN}^{\mathrm{eff}}(E)
=
\int
\sigma_{NN}(p_{\mathrm{rel}})
D(p_{\mathrm{rel}})\,
dp_{\mathrm{rel}},
\label{eq:sigma_eff}
\end{equation}
where $D(p_{\mathrm{rel}})$ represents the relative-momentum distribution. For axially deformed projectiles, the cross section must be averaged over intrinsic orientations,
\begin{equation}
\langle \sigma_I \rangle
=
\frac{1}{4\pi}
\int
\sigma_I(\Omega)\,
d\Omega.
\label{eq:orientation_sigma}
\end{equation}
In the present study, orientation averaging is performed for $^{29}$Ne and $^{31}$Ne.

The interaction cross section is primarily sensitive to the surface and low-density tail of the matter distribution.
Therefore, nuclei with extended low-density tails, such as halo candidates, are expected to exhibit enhanced
$\sigma_I$.
As shown in Sec.~III, $^{31}$Ne possesses a significantly enhanced Woods--Saxon diffuseness parameter,
indicating a slowly decaying radial density profile.
The enhanced $\sigma_I$ observed in Fig.~\ref{fig:sigma_combined} is therefore a direct reaction/interaction
signature of the same extended neutron density.

\subsection{Energy correction to $E_{\rm ref}=240$ MeV/nucleon and multipole-truncation test}
\label{subsec:sigma240_energycorr_lmax}

\begin{figure*}[t]
  \centering
  \begin{minipage}{0.48\textwidth}
    \centering
    \includegraphics[width=\linewidth]{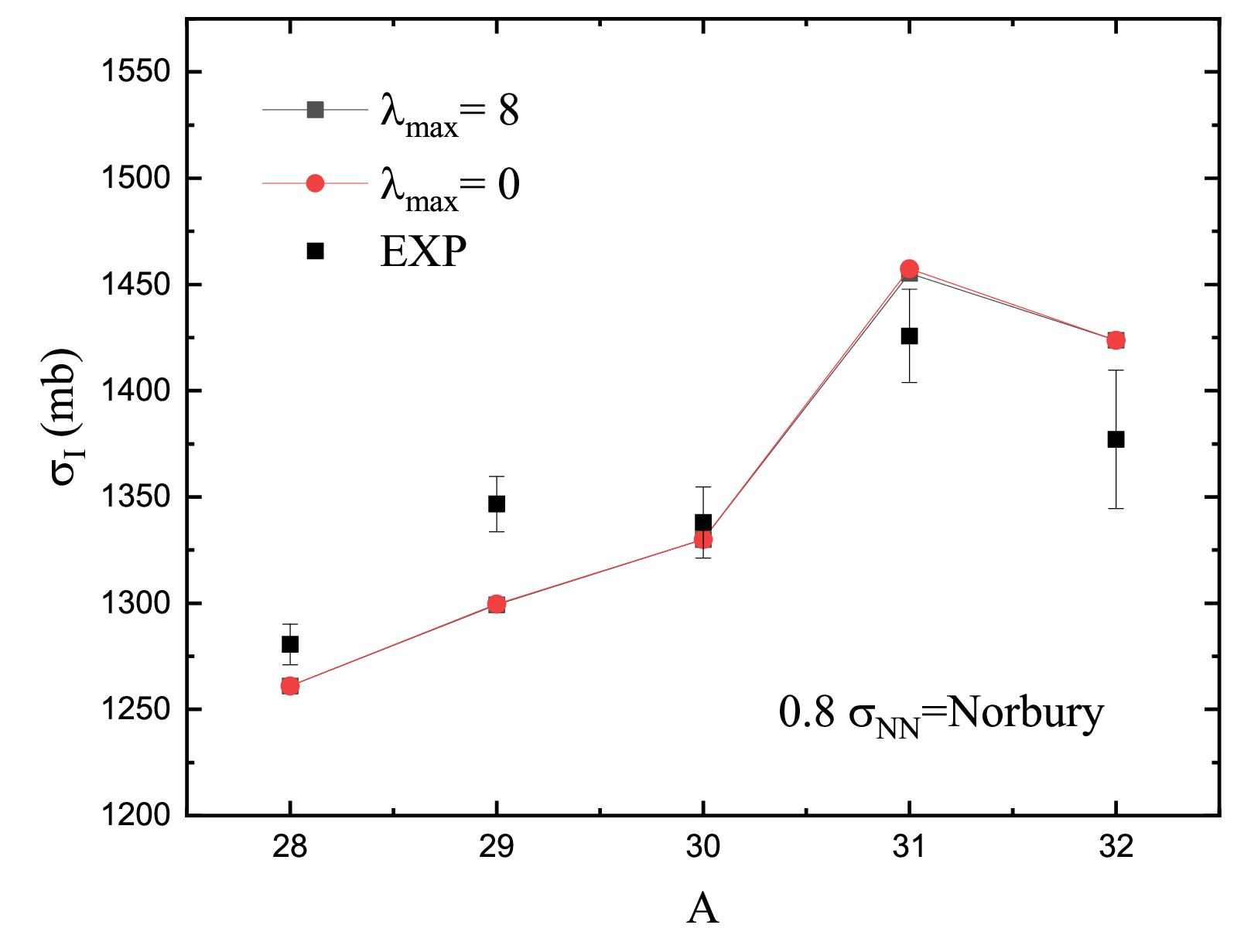}
    \vspace{-2mm}
    \textbf{(a)}
  \end{minipage}
  \hfill
  \begin{minipage}{0.48\textwidth}
    \centering
    \includegraphics[width=\linewidth]{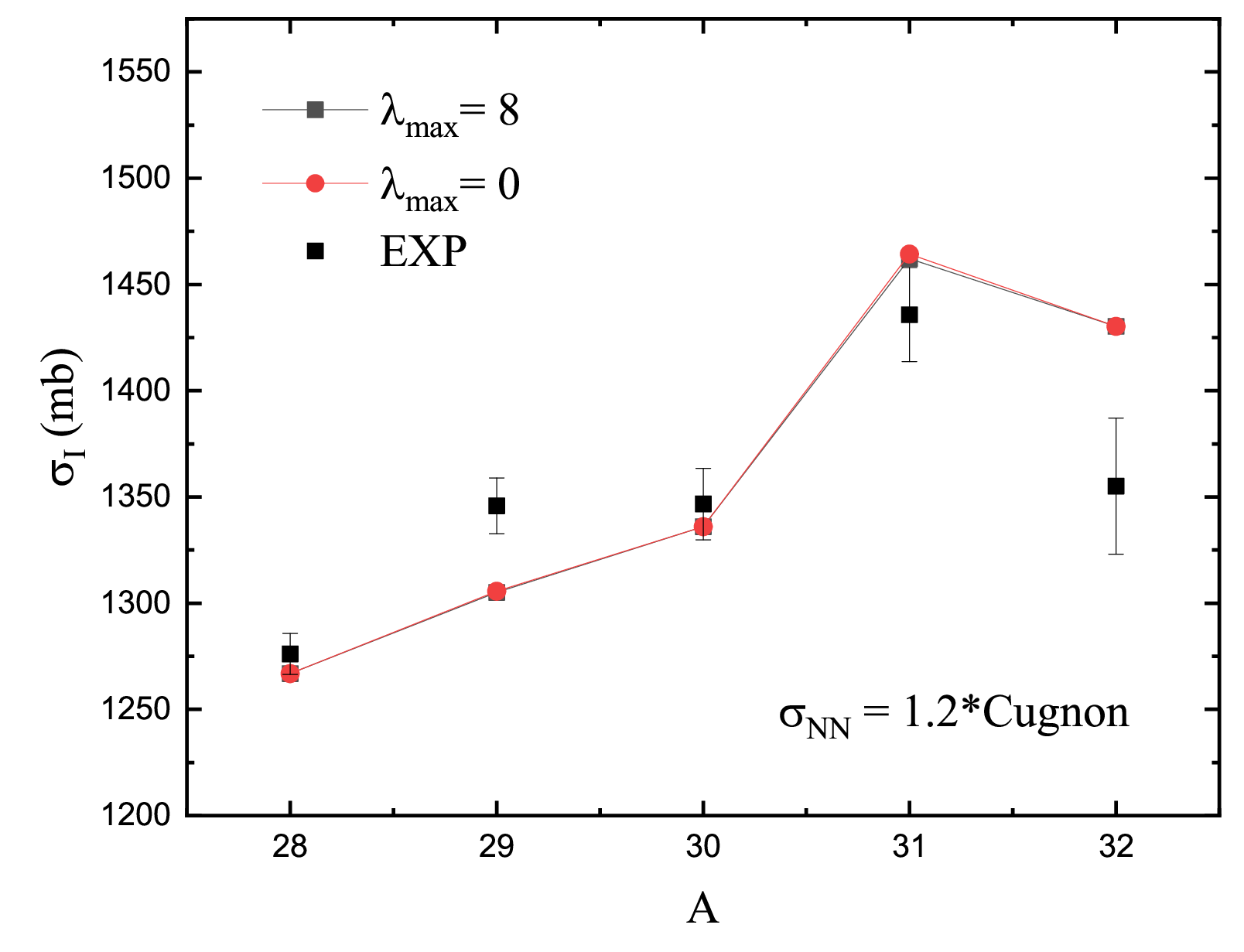}
    \vspace{-2mm}
    \textbf{(b)}
  \end{minipage}
  \caption{(Color online)
  Interaction cross sections $\sigma_I$ for
  $^{28\text{--}32}$Ne on a $^{12}$C target at the reference energy
  $E_{\rm ref}=240$ MeV/nucleon calculated within the Glauber MOL[FM] framework
  including the Coulomb correction~\cite{AbuIbrahim2000,Takechi2009}.
  The projectile density is treated with two multipole truncations:
  $\lambda_{\max}=8$ and $\lambda_{\max}=0$.
  Panel (a) uses the Norbury(2008) NN parametrization~\cite{Norbury2008}
  (with the adopted global rescaling, $0.8\times$Norbury),
  while panel (b) uses the Cugnon(1996) NN parametrization~\cite{Cugnon1996}
  (with the adopted global rescaling, $1.2\times$Cugnon).
  Experimental points from Ref.~\cite{Takechi2012} are converted to equivalent values
  at $E_{\rm ref}=240$ MeV/nucleon using the energy-correction procedure in
  Eq.~(\ref{eq:energy_correction_240}), evaluated consistently for each NN-input prescription.}
  \label{fig:sigma_240_lmax_combined}
\end{figure*}

To perform a controlled comparison at a common reference energy,
we evaluate theoretical cross sections at $E_{\rm ref}=240$ MeV/nucleon within the same
Glauber MOL[FM] prescription~\cite{AbuIbrahim2000,Takechi2009}, including the Coulomb correction.

\paragraph{Coulomb correction.}
In the present implementation, the Coulomb deflection is incorporated at the level of the
impact-parameter integration through the Jacobian factor associated with the transformation from the
asymptotic impact parameter to the distance of closest approach along a Rutherford trajectory.
The correction factor can be expressed as
\begin{equation}
C(b)=\max\!\left[0,\,1-\frac{a}{b}\right],
\qquad
a=\frac{Z_P Z_T e^2}{2E_{\rm cm}},
\label{eq:coulomb_factor}
\end{equation}
where $E_{\rm cm}$ is approximated using masses proportional to $A$.
In the numerical evaluation, the interaction probability in the $b$-integration is multiplied by $C(b)$.

\paragraph{Energy correction of experimental points.}
Because the measured interaction cross sections are reported at slightly different beam energies $E_{\rm exp}$~\cite{Takechi2012},
the directly measured values are converted to equivalent values at $E_{\rm ref}=240$ MeV/nucleon using the
energy dependence predicted by the same Glauber MOL[FM] calculation,
\begin{equation}
\sigma_I^{\rm eq}(E_{\rm ref})
=
\sigma_I^{\rm exp}(E_{\rm exp})
\,
\frac{\sigma_I^{\rm cal}(E_{\rm ref})}{\sigma_I^{\rm cal}(E_{\rm exp})}.
\label{eq:energy_correction_240}
\end{equation}
Importantly, the correction factor is evaluated consistently for each NN-input prescription,
i.e., the Norbury-based $\sigma_I^{\rm cal}$ is used for the Norbury panel and the Cugnon-based
$\sigma_I^{\rm cal}$ is used for the Cugnon panel. \red{This model ratio is used only to place the experimental points on a common reference-energy axis for comparison; it does not alter the measured isotopic ordering and is applied separately for each NN prescription.}

\paragraph{Multipole truncation and orientation averaging.}
To quantify the deformation-induced anisotropy in the interaction observable, we perform a diagnostic
multipole-truncation test by comparing the results obtained with
(i) $\lambda_{\max}=8$ and (ii) $\lambda_{\max}=0$ (monopole component only).
Orientation averaging is performed for the deformed isotopes $^{29}$Ne and $^{31}$Ne.
As shown in Fig.~\ref{fig:sigma_240_lmax_combined},
the difference between $\lambda_{\max}=8$ and $\lambda_{\max}=0$ remains small after orientation averaging,
while the $^{31}$Ne enhancement persists, indicating that the anomaly is dominated by the extended
low-density tail of the monopole density rather than by geometric deformation effects.

\subsection{Peripheral origin of the $^{31}$Ne enhancement: impact-parameter decomposition}
\label{subsec:bprofile}

\begin{figure*}[t]
  \centering
  \begin{minipage}{0.49\textwidth}
    \centering
    \includegraphics[width=\linewidth]{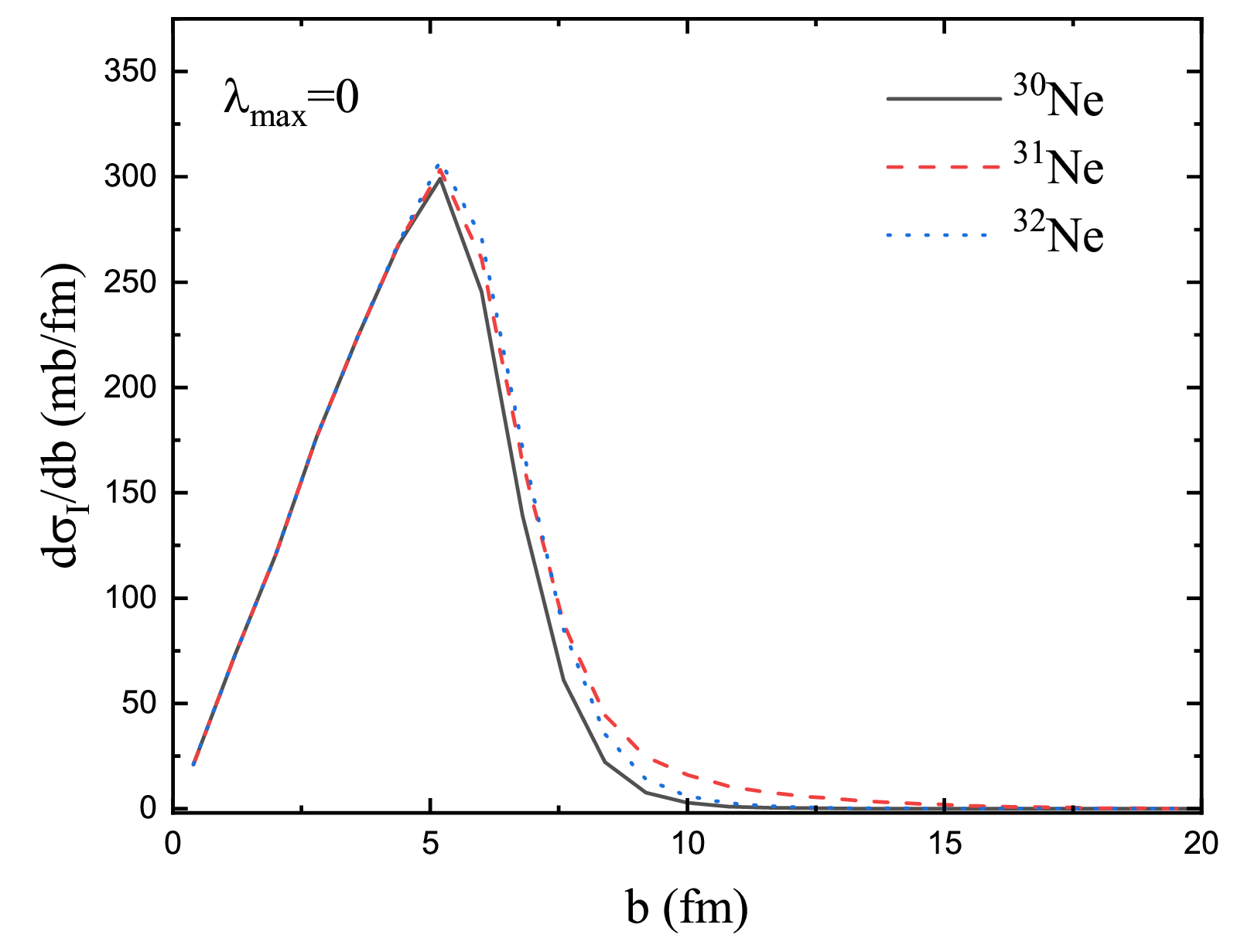}
    \vspace{-2mm}
    \textbf{(a)}
  \end{minipage}
  \hfill
  \begin{minipage}{0.49\textwidth}
    \centering
    \includegraphics[width=\linewidth]{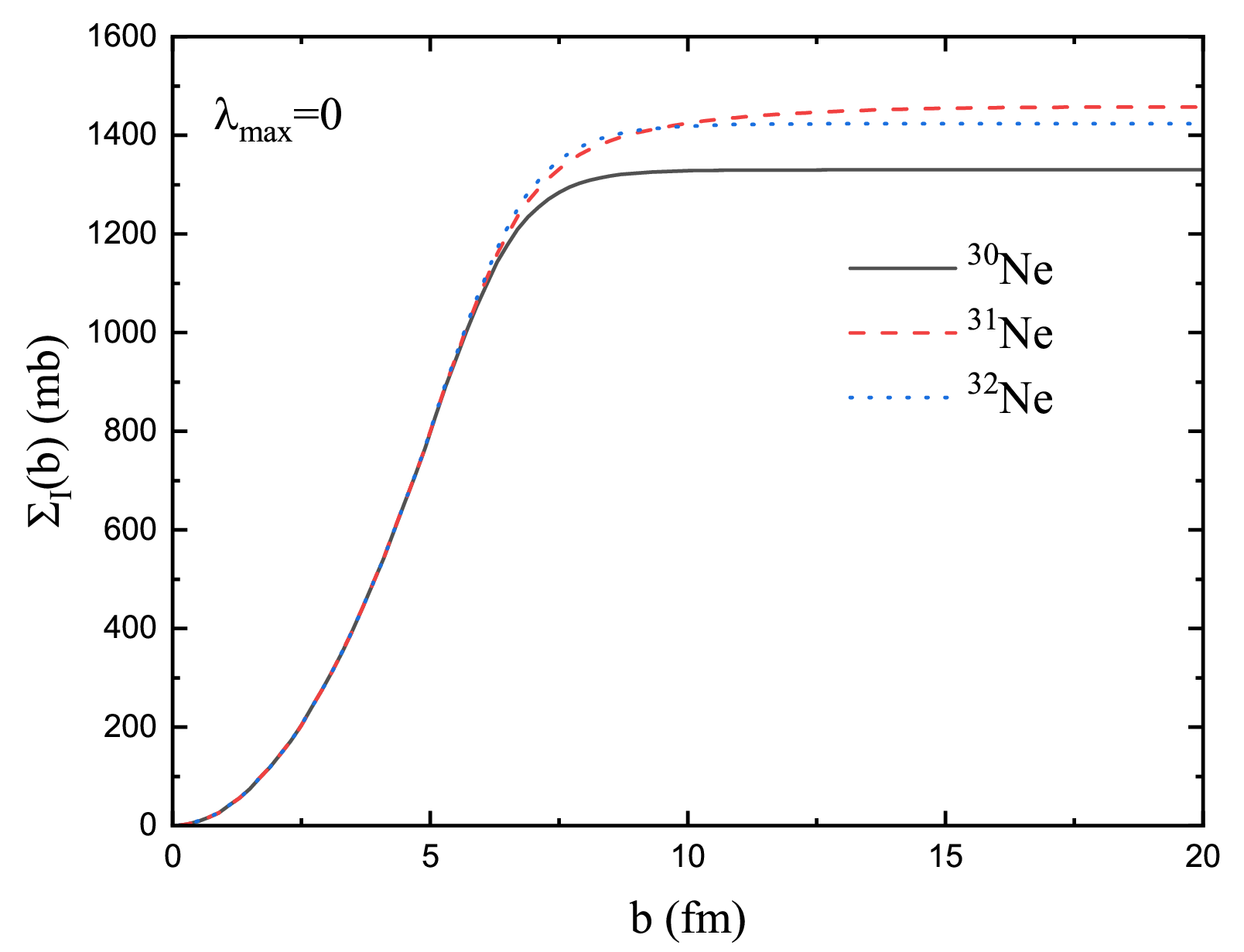}
    \vspace{-2mm}
    \textbf{(b)}
  \end{minipage}
  \caption{(Color online)
  Impact-parameter decomposition of the interaction cross section for
  $^{30,31,32}$Ne+$^{12}$C at $E_{\rm ref}=240$ MeV/nucleon calculated within the
  Glauber MOL[FM] framework~\cite{AbuIbrahim2000,Takechi2009} including the Coulomb correction.
  The Norbury(2008) NN parametrization~\cite{Norbury2008} with the adopted global rescaling
  ($0.8\times$Norbury) is used.
  The projectile density is truncated to the monopole component
  ($\lambda_{\max}=0$).
  Panel (a) shows the differential contribution $d\sigma_I/db$ obtained by radial binning
  of the two-dimensional impact-parameter integrand, and panel (b) shows the cumulative cross section
  $\Sigma_I(b)$.
  The enhancement of $^{31}$Ne is accumulated predominantly in the peripheral region
  ($b\gtrsim 8$ fm), indicating a tail-dominated origin of the interaction anomaly.
  (The full multipole result up to $\lambda_{\max}=8$ is nearly indistinguishable and is omitted.)}
  \label{fig:bprofile_240}
\end{figure*}

To identify which impact-parameter region is responsible for the enhanced
$\sigma_I$ of $^{31}$Ne, we decompose the two-dimensional
impact-parameter integral at
$E_{\rm ref}=240$ MeV/nucleon within the same Glauber MOL[FM] prescription
used in the previous subsection (including the Coulomb correction). For numerical convenience, we define the attenuation (opacity) function
\begin{equation}
\chi_I(\mathbf b)\equiv 2\,\mathrm{Im}\,\chi_{\mathrm{MOL}}(\mathbf b),
\end{equation}
so that the integrand of Eq.~(\ref{eq:sigmaI_final}) is written as
$1-\exp\{-\chi_I(\mathbf b)\}$.
On the transverse grid, the interaction cross section is evaluated as
\begin{equation}
\sigma_I \;=\; \int d^2b \; C(b)\,\Big[1-\exp\{-\chi_I(\mathbf b)\}\Big],
\end{equation}
where $C(b)$ is the Coulomb correction factor and $\chi_I(\mathbf b)$ is obtained from the MOL phase
in Eq.~(\ref{eq:chi_mol}).

We define the contribution in a radial bin $[b_k,b_{k+1})$ as
\begin{equation}
\Delta\sigma_I(b_k)
=
\int_{b_k}^{b_{k+1}} d^2b \; C(b)\,\Big[1-\exp\{-\chi_I(\mathbf b)\}\Big]
\;\approx\;
\sum_{\mathbf b_{ij}\in[b_k,b_{k+1})}
C(b_{ij})\Big[1-e^{-\chi_{I,ij}}\Big]\,(\Delta x)^2 ,
\end{equation}
and the corresponding differential and cumulative distributions as
\begin{equation}
\left.\frac{d\sigma_I}{db}\right|_{b_k}\approx \frac{\Delta\sigma_I(b_k)}{\Delta b},
\qquad
\Sigma_I(b)=\int_0^{b}\frac{d\sigma_I}{db'}\,db'.
\end{equation}

Figure~\ref{fig:bprofile_240} shows $d\sigma_I/db$ and $\Sigma_I(b)$ for
$^{30,31,32}$Ne+$^{12}$C using the Norbury(2008) NN input~\cite{Norbury2008}
with the adopted global rescaling ($0.8\times$Norbury).
For clarity we display the monopole-truncated projectile density
($\lambda_{\max}=0$), which isolates the tail contribution from deformation-induced anisotropy.
We verified that the full multipole calculation up to $\lambda_{\max}=8$
gives an almost indistinguishable result at the level of both
$d\sigma_I/db$ and $\Sigma_I(b)$ and therefore is not shown.

The peak region around $b\sim 4$--$6$ fm is similar among the isotopes,
whereas $^{31}$Ne exhibits a pronounced enhancement in the peripheral region
($b\gtrsim 8$ fm).
In the present calculation, the contribution from $b\ge 10$ fm amounts to
$\sim 34$ mb for $^{31}$Ne, compared with $\sim 2$ mb ($^{30}$Ne) and
$\sim 6$ mb ($^{32}$Ne), demonstrating that the $^{31}$Ne anomaly in
$\sigma_I$ is generated predominantly by the extended low-density tail.

\section{Summary and Outlook}

We have carried out a systematic study of $^{28\text{--}32}$Ne by combining microscopic DRHBc densities, phenomenological Woods--Saxon parametrization, Helm-model form-factor analysis, and Glauber reaction cross section calculations. The DRHBc densities reveal a pronounced neutron extension in $^{31}$Ne, while $^{29}$Ne does not exhibit a comparable tail enhancement and $^{32}$Ne remains intermediate. \red{Across these complementary analyses, $^{31}$Ne stands out most clearly through its anomalously large surface diffuseness, whereas the reduction of the fitted radius parameter is interpreted only as a secondary consequence of the tail-sensitive WS fit.}

The Helm-model analysis shows that deformation contributes to geometric smearing but does not fully account for the observed spatial extension in $^{31}$Ne. In the reaction sector, the calculated interaction cross sections reproduce the overall isotopic trend and maintain the relative enhancement of $^{31}$Ne across the NN prescriptions considered here. \red{The impact-parameter decomposition further shows that this enhancement is accumulated predominantly in the peripheral region, consistent with a tail-dominated origin of the anomaly.}

Taken together, these analyses consistently identify $^{31}$Ne as the most prominent halo candidate within the $^{28\text{--}32}$Ne isotopic chain, while $^{32}$Ne exhibits intermediate features that may reflect a borderline halo or an enhanced neutron skin, and $^{29}$Ne shows no clear halo signature. \red{The halo character of $^{31}$Ne is supported by four mutually consistent signatures: microscopic density extension, enhanced Woods--Saxon diffuseness, Helm-model mismatch, and a robust relative enhancement of the reaction cross section.} The framework developed here can be extended to neighboring isotopic chains such as Mg and Al, to more detailed continuum and single-particle analyses for $^{32}$Ne, and to reaction models beyond the optical-limit treatment. More precise measurements of interaction cross sections, momentum distributions, and Coulomb dissociation observables would provide additional constraints on the interpretation of $^{31}$Ne and clarify the nature of $^{32}$Ne.

\section*{Acknowledgement}
This work was supported by the National Research Foundation of Korea (Grant Nos. RS-2025-00513410 and RS-2024-00460031).  The work of MKC is supported by the National Research Foundation of Korea (Grant Nos. RS-2021-NR060129 and RS-2025-16071941). 
%


\end{document}